\documentclass[aps,prappl,reprint,superscriptaddress,showpacs,floatfix,longbibliography]{revtex4-1}%

\usepackage[T1]{fontenc}
\usepackage{mathptmx}

\usepackage{graphicx}
\usepackage{amssymb} 
\usepackage{dcolumn}
\usepackage{bm}
\usepackage[mathlines]{lineno}
\usepackage[colorlinks,linkcolor=blue,anchorcolor=blue,citecolor=blue,urlcolor=blue]{hyperref}
\usepackage{multirow}
\usepackage{color}

\makeatletter
\renewcommand*{\@fnsymbol}[1]{\ensuremath{\ifcase#1\or \dagger\or *\or \ddagger\or
\mathsection\or \mathparagraph\or \|\or **\or \dagger\dagger \or
\ddagger\ddagger \else\@ctrerr\fi}} \makeatother

\usepackage{color}%

\begin{document}
%
\title{Kirigami-inspired Thermal Regulator}


\author{Hongyi Ouyang\footnotemark[1]}
\thanks{These authors contributed equally to this work.}
\affiliation{State Key Laboratory for Mechanical Behavior of Materials, Xi'an Jiaotong University, Xi'an 710049, China}

\author{Yuanqing Gu}
\thanks{These authors contributed equally to this work.}
\affiliation{Key Laboratory of Mechanism and Equipment Design of Ministry of Education, Tianjin University, Tianjin, 300350, China}
\affiliation{School of Mechanical Engineering, Tianjin University, Tianjin, 300350, China}

\author{Zhibin Gao}
\email[E-mail: ]{zhibin.gao@xjtu.edu.cn}
\affiliation{State Key Laboratory for Mechanical Behavior of Materials, Xi'an Jiaotong University, Xi'an 710049, China}

\author{Lei Hu}
\affiliation{State Key Laboratory for Mechanical Behavior of Materials, Xi'an Jiaotong University, Xi'an 710049, China}

\author{Zhen Zhang}
\affiliation{State Key Laboratory for Mechanical Behavior of Materials, Xi'an Jiaotong University, Xi'an 710049, China}

\author{Jie Ren}
\affiliation{Center for Phononics and Thermal Energy Science, China-EU Joint Center for Nanophononics, Shanghai 
                Key Laboratory of Special Artificial Microstructure Materials and Technology, School of Physics Sciences 
                and Engineering, Tongji University, Shanghai 200092, China}


\author{Baowen Li}
\affiliation{Department of Materials Science and Engineering, Department of Physics. Southern University of Science 
                and Technology, Shenzhen, 518055, PR China. International Quantum Academy, Shenzhen 518048, PR 
                China. Paul M. Rady Department of Mechanical Engineering and Department of Physics, University of 
                Colorado, Boulder, Colorado 80305-0427, USA}                

\author{Jun Sun}
\affiliation{State Key Laboratory for Mechanical Behavior of Materials, Xi'an Jiaotong University, Xi'an 710049, China}                
                
\author{Yan Chen}
\email[E-mail: ]{yan\_chen@tju.edu.cn}
\affiliation{Key Laboratory of Mechanism and Equipment Design of Ministry of Education, Tianjin University, Tianjin, 300350, China}
\affiliation{School of Mechanical Engineering, Tianjin University, Tianjin, 300350, China}

\author{Xiangdong Ding}
\email[E-mail: ]{dingxd@mail.xjtu.edu.cn}
\affiliation{State Key Laboratory for Mechanical Behavior of Materials, Xi'an Jiaotong University, Xi'an 710049, China}

\date{\today}
\begin{abstract}
One of the current challenges in nanoscience is tailoring phononic devices such 
as thermal regulators and thermal computing. This has long been a rather elusive task 
because the thermal switching ratio is not as high as electronic analogs. %
Mapping from a topological kirigami assembly, nitrogen-doped porous 
graphene (NPG) metamaterials on the nanoscale are inversely designed with a 
thermal switching ratio of 27.79 which is more than double the value of previous 
work. We trace this behavior to the chiral folding-unfolding deformation, resulting 
in a metal-insulator transition. %
%
This study provides a nano-material design paradigm to bridge the gap between kinematics and functional metamaterials that motivates the development of 
high-performance thermal regulators. %
\end{abstract}



\maketitle





Thermal regulation is the most critical technology challenge for modern 
electronics~\cite{waldrop2016chips, Philip2012computer},  
thermoelectricity~\cite{zhao2014ultralow}, %
thermal cooling~\cite{jiang2022inhibiting, lindsay2013first}, interfacial 
thermal resistance~\cite{chen2022interfacial} and 
phononics~\cite{li2012colloquium, gu2018colloquium}. Similar to 
electrical switches, thermal regulators whose thermal transport in two 
states or directions is relatively different from each 
other~\cite{li2004thermal, wang2007thermal, chang2006solid} and play 
important roles in thermal storage, thermal management and solid-state 
thermal circuits~\cite{martinez2015rectification, wang2008thermal, wehmeyer2017thermal, li2021transforming}. %
To regulate the thermal conductivity $\kappa$, researchers focused on the 
modulation between two states and mechanisms that facilitate large 
thermal switching ratio $R$ between on-state high thermal conductivity 
${\kappa_{\rm on}}$ and off-state low thermal conductivity ${\kappa_{\rm off}}$. %

Several materials have been unveiled with thermal switching behavior 
under different kinds of incentives~\cite{jimenez2021acoustic}, such as shape memory 
alloy~\cite{hao2018efficient,li2015temperature}, shape-dependent 
nanoislands~\cite{bode2004shape},monolayer lateral heterojunction~\cite{zhang2022simultaneous},
multi-body effect in radiation~\cite{thompson2020nanoscale}, strain in ferroelectric 
domain~\cite{langenberg2019ferroelectric} and in compressible 
graphene~\cite{du2021wide}, thermophile 
proteins~\cite{nonoyama2020instant}, plasmon 
resonators~\cite{ilic2018active}, local effect in 
polymer~\cite{azulay2003electrical}, electric 
field~\cite{ihlefeld2015room,liu2019electric}, magnetic 
field~\cite{shin2016thermally}, temperature-induced phase 
transition~\cite{shrestha2019high},  temperature-dependent transformation thermotics,
element 
substitution~\cite{aryana2021suppressed}, electrochemical 
reaction~\cite{cho2014electrochemically, lu2020bi}, optical 
excitation~\cite{shin2019light} and even hydration in bio-inspired 
materials~\cite{tomko2018tunable}. Although these studies have
reported up to an order of magnitude in switching ratios 
and corresponding mechanisms, how to further augment $R$
for high-performance thermal regulators is still a conundrum
and limits their applications. %

\begin{figure}
\includegraphics[width=1.0\columnwidth]{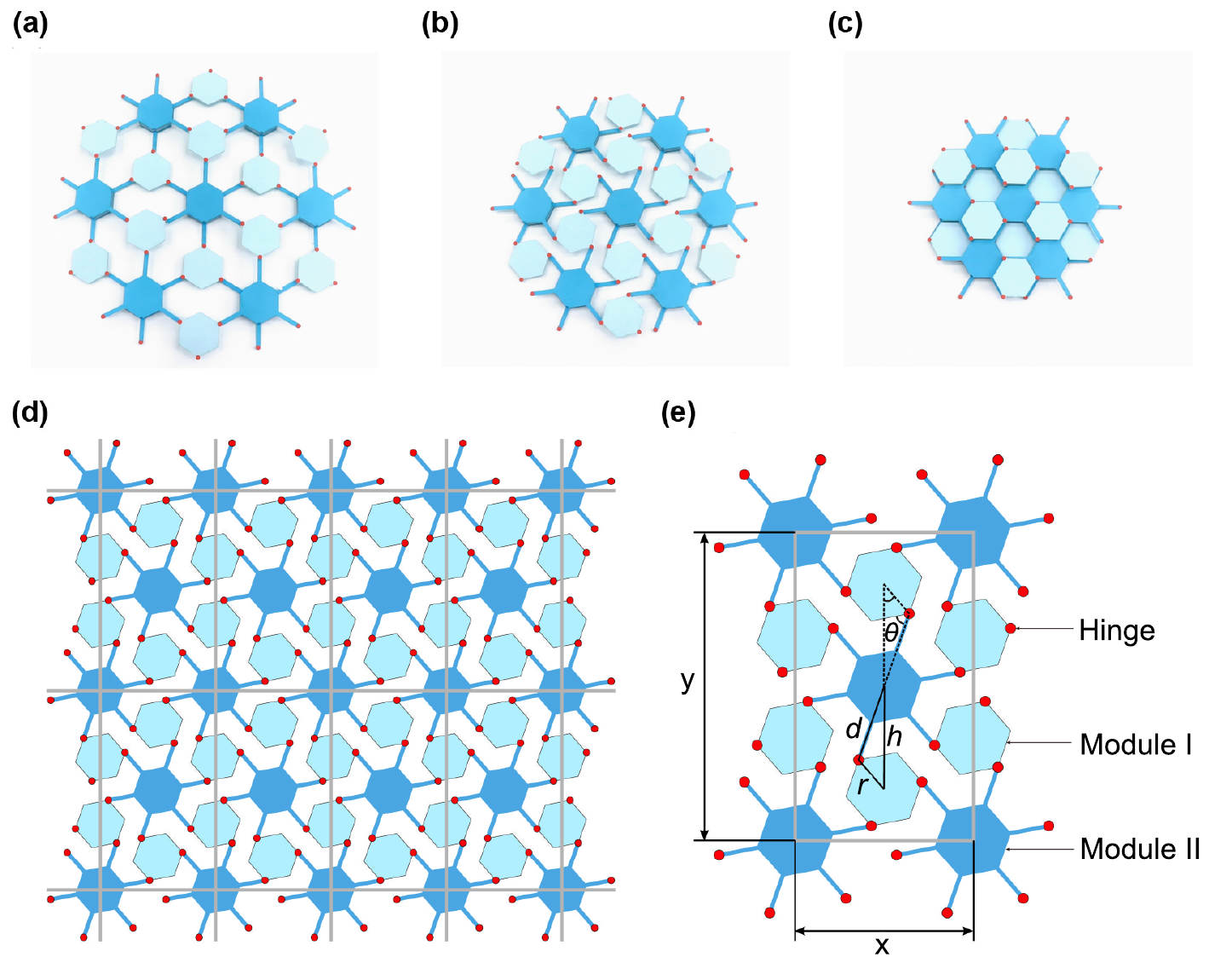}
\caption{
Kirigami assembly consisting of regular hexagons. %
(a-c) Snapshots of the physical prototype in unfolding, intermediate, and fully closed states, %
respectively. Hinge points are shown by red dots. (d) Periodic tessellation %
is displayed by grey rectangles. (e) Geometric %
parameters are defined in the unit cell. %
\label{fig1}}
\end{figure}

In this Letter, we design a 2D honeycomb kirigami assembly using metamaterial
and map it on porous graphene.
Chirality is induced by nitrogen substitution, realizing asymmetrical folding-unfolding deformation. %
We find NPG can exhibit an ultrahigh $R$ of 27.78 
under only 1\% strain. With this small deformation, the obvious change was found in 
the band structure, resulting in an unusual metal-insulator transition and 
enhanced optical absorption in NPG. Our work offers a  nano-material design 
paradigm mapping from topological assembly on metamaterials, showing potential 
high-performance thermal regulators. %
The 2D honeycomb kirigami assembly based on the nanoporous graphene
with the folding-unfolding deformation, shown in Fig.~\ref{fig1}.  It consists
of two modules, one of which is module I having a hexagon with 3 hinge points (light blue). Another 
one is module II, a similar hexagon with 6 arms (dark blue). The endpoints of the arms are 
connected with module I by hinges.
The kirigami model enables the entire assembly with one degree-of-freedom motion and avoids physical interference. Snapshots of kirigami assembly during deformation are 
shown in Fig.~\ref{fig1}(a-c), and the tessellation is presented in Fig.~\ref{fig1}(d). The 
folding-unfolding deformation for the kirigami assembly is depicted in Video 1. %

\begin{figure}[bp]
\includegraphics[width=1.0\columnwidth]{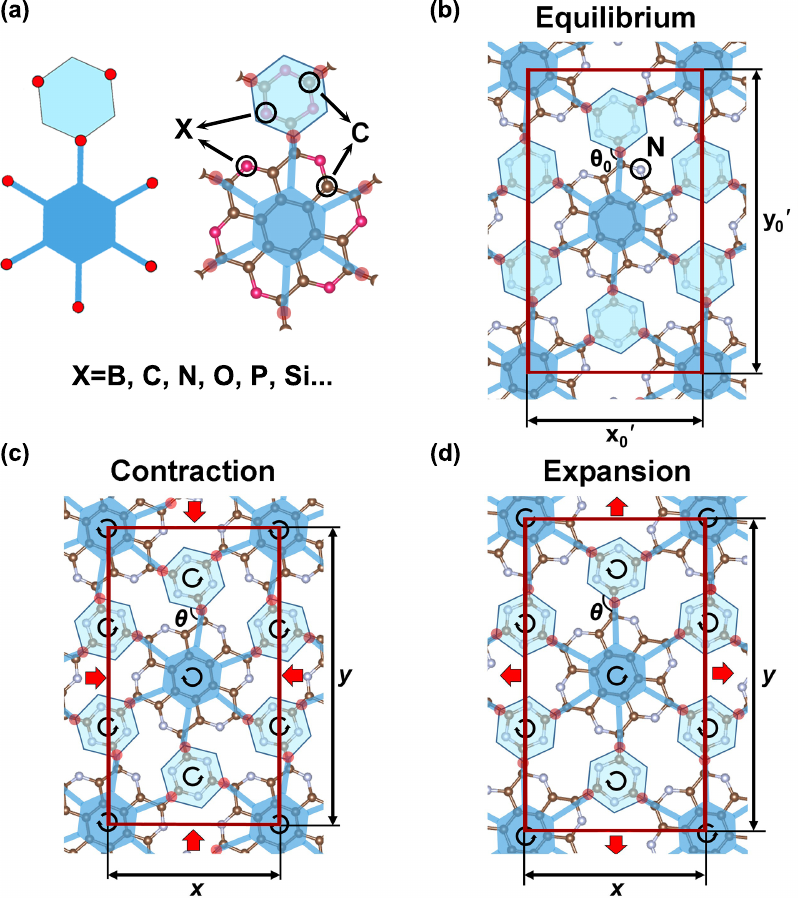}
\caption{%
Mapping of kirigami assembly to metamaterials. (a) Module I and II are replaced by 
one honeycomb and a cluster including seven honeycombs. %
The element of pink atoms labeled by X is to be decided shown in Supplemental Material. %
Snapshots of NPG in (b) equilibrium, (c) contraction, and (d) expansion states %
during deformation. 
Black arrows indicate the rotation of modules. %
The brown and white balls are C and N, respectively.
%
\label{fig2}}
\end{figure}

Geometric parameters of the kirigami assembly are illustrated in Fig.~\ref{fig1}(e). $r$ and $d$ 
are the distance between center modules and hinge points in modules I and II, respectively. $h$ 
is the distance between the center of two modules and $\theta$ is the angle between module I 
and the adjacent arm. A periodical rectangular unit cell was depicted in the kirigami assembly, 
which is more easily to determine the strain in the Cartesian coordinate system. Then we 
investigate the deformation behavior of the assembly, %
\begin{eqnarray}
&h=\sqrt{d^{2}+r^{2}-2dr \cdot \cos(\theta+\pi/3)}&,
\end{eqnarray}
%
\begin{eqnarray}
&x=\sqrt3h{\rm ,}~~~~~~~~~y=3h&.
\end{eqnarray}

%


\begin{figure*} 
\includegraphics[width=2.0\columnwidth]{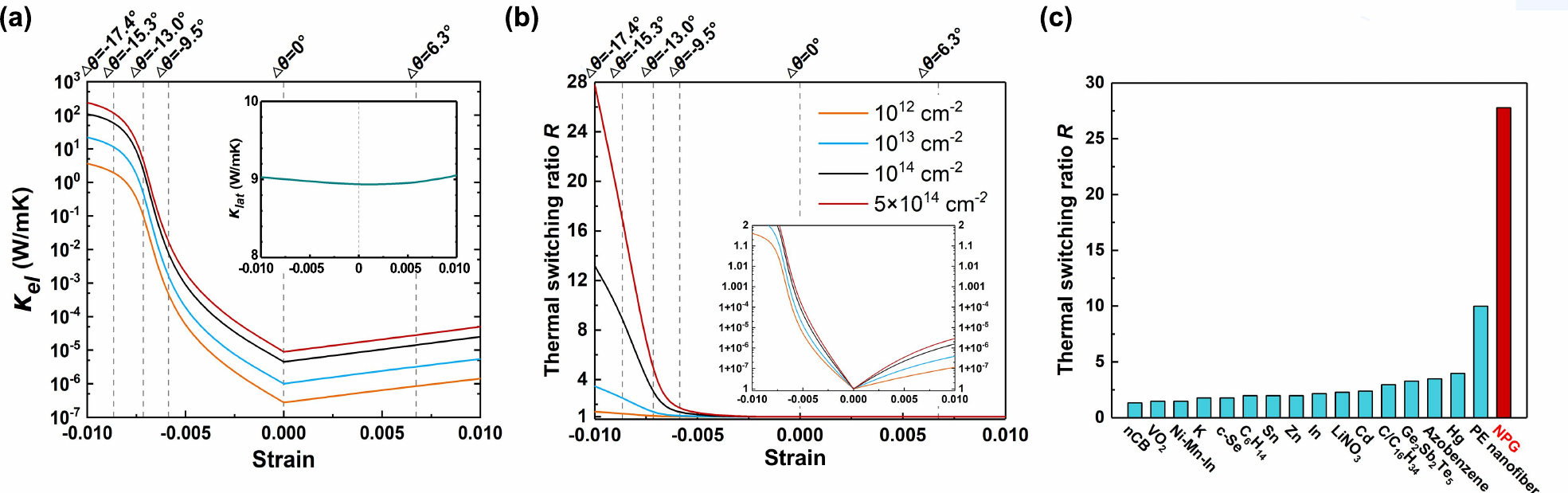}
\caption{%
(a) ${\kappa}_{el}$ and ${\kappa}_{lat}$ (inset) of NPG as functions of strain. %
(b) Thermal switching ratio $R$ versus strain under different %
hole carrier concentrations. The zoom feature of $R$ between 1 and 2 is shown in 
the inset. A maximum doping level of $5 \times 10^{-14}~{\rm cm^{-2}}$  for both 
electrons and holes has been reached by electrical gating~\cite{efetov2010controlling} 
and ionic liquid injection~\cite{chuang2014high}. %
(c) $R$ of NPG compared with reported thermal regulators. %
nCB~\cite{marinelli1998thermal}; ${\rm VO_{2}}$~\cite{oh2010thermal}; %
K, Zn, Cd, and Hg~\cite{ho1972thermal}; c-Se~\cite{abdullaev1966thermal}; %
Ni-Mn-In alloy~\cite{zheng2019high}; ${\rm C_{6}H_{14}}$~\cite{konstantinov2011isochoric}; %
${\rm Sn, In, and~LiNO_{3}}$~\cite{kim2016thermal}; %
${\rm C/C_{16}H_{34}}$ composite~\cite{zheng2011reversible}; %
${\rm Ge_{2}Sb_{2}Te_{5}}$~\cite{crespi2014electrical}; %
azobenzene polymers~\cite{shin2019light}; PE nanofiber~\cite{shrestha2019high}. %
%
\label{fig3}}
\end{figure*}




In the following, we map the kirigami assembly on 2D mechanical 
metamaterials. The point-to-point rigid rod connections in kirigami assembly are replaced
by chemical bonds.  
According to the spatial linkage kinematics about the relationship between cyclohexane 
molecular and octahedral Bricard 6R linkage~\cite{baker1986limiting}, we call the bond 
connecting different modules as ``hinge bond'' and treat the midpoints of the ``hinge 
bond'' as hinges. Therefore, two components were designed to 
represent two modules in the assembly, as shown in Fig.~\ref{fig2}(a). To increase the 
stability of the structure, carbocyclic six-membered rings are kept. 
What's more, chirality can be induced by elemental substitution, resulting in 
the asymmetrical folding-unfolding deformation in metamaterials.

Since the potential element is selected, we set criteria to guide the element selection 
labeled X shown in Fig.~\ref{fig2}(a), which include (i) obvious distortion out of plane 
should be avoided that 
will preclude the deformation, (ii) the connection between two types of modules 
should be only one chemical bond to ensure the flexibility of whole structure, %
(iii) the electronic band structure in the equilibrium state should be an insulator to 
realize the metal-insulator transition by small deformation. The whole detail of the selection process with 
element X is discussed in Supplemental Material~\cite{tsetseris2014substitutional}. Consequently, the nitrogen element
is the best one of X and we focus on the nitrogen porous graphene, named NPG. %
We acquired the optimized metamaterials with different lattice 
constants and $\theta$ by Density Function Theory (DFT). The details of calculation are shown in the Supporting Information~\cite{blochl1994projector, kresse1999ultrasoft, perdew1996generalized, kresse1996efficiency, kresse1996efficient, monkhorst1976special}.
Interestingly, NPG in Fig.~\ref{fig2}(b-d) and Video 2 performs a similar 
folding-unfolding deformation behavior after contraction or expansion, corresponding to the
kirigami assembly in Fig.~\ref{fig1}(a-c). The most stable structure has a 
${\rm \theta_0}$ of $110.73^\circ$ which is slightly different from the assembly with $120^\circ$. 

We unexpectedly find that NPG has a property of chirality which means the rotation in 
clockwise and anti-clockwise exhibit different deformation behaviors~\cite{coulais2018multi}. %
More mechanical property of NPG is discussed in the Supplemental Material. %


Now we address the crucial issue, namely, 
deformation effect on the electronic and thermal properties. Thermal conductivity 
can be divided into two parts. Based on \textit{ab initio} and non-equilibrium molecular 
dynamics simulation, we 
can obtain $\kappa_{lat}$ and $\kappa_{el}$ by Fourier law and 
Wiedemann-Franz (WF) law, %
%
%
\begin{eqnarray}
&\displaystyle\kappa_{lat}=-\frac{J}{{\nabla}T}, ~~~~~~\displaystyle\kappa_{el}={\rm L_0}{\sigma}T&.  %
\end{eqnarray}
%
%
%

\noindent
where $J$ is the heat flux of energy transport per unit time across the unit area that 
is defined as $S=W \times d$ in which $W$ is the width of the 2D sheet and 
$d$ is the inter-layer distance for the corresponding bulk 
material ($d=3.355$~\AA~for NPG). $\nabla T$ is the temperature gradient along the
heat transport direction. $\sigma$ is the electric conductivity and ${\rm L_0}$ is the 
Lorenz constant number of $2.44\times 10^{-8}$~W~${\rm\Omega~K^{-2}}$. The discussion about
why WF law is suited to NPG~\cite{crossno2016observation, bardeen1950deformation, xi2012first, cai2014polarity, yang2016two, zhu2017multivalency}, detials of calculation~\cite{plimpton1995fast, kinaci2012thermal} and length dependence of $\kappa_{lat}$~\cite{xu2014length, wang2017experimental, lepri1997heat, gao2016heat} are shown in Supplemental Material.


We find that the change of $\kappa_{lat}$ is not obvious as a function of 
strain, while $\kappa_{el}$ increases by 7 orders of magnitude after 
contraction of only 1\% strain shown in Fig.~\ref{fig3}(a). This discovery will 
result in a dominating role of ${\kappa}_{el}$ compared with ${\kappa}_{lat}$. %
$\kappa_{lat}$ has a value of 8.92~W/mK in equilibrium state and fluctuates 
slightly  to 9.05 W/mK with 1\% compressive strain. $\kappa_{el}$ 
of different doping carriers and concentrations are shown in Fig. S12. %

\begin{figure*}
\includegraphics[width=2.0\columnwidth]{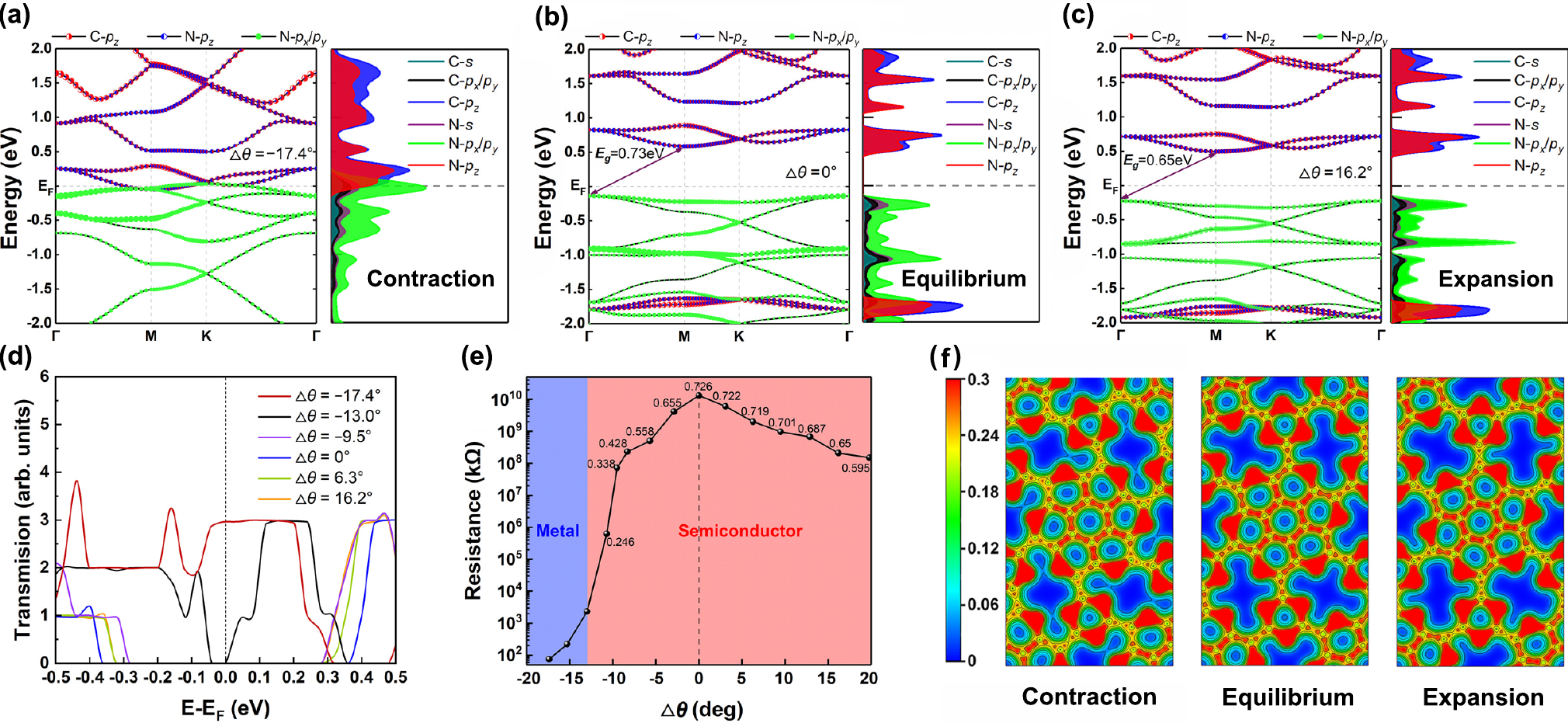}
\caption{
(a-c) Projected band structures and PDOS %
of NPG in three deformation states. %
(d) Electronic transmission spectrum and (e) electrical resistance of NPG device with different $\Delta\theta$.
Fermi level was regarded to be zero for each 
state to compare the transfer ability of electrons between electrodes among all states. %
The number in (e) stands for the band gap $E_g$ in the eV unit. %
(f) The isosurface plot of the charge density of NPG 
and the isosurface value is taken as 0.06 e/${\rm Bohr^{3}}$. 
\label{fig4}}
\end{figure*}

Thermal switching ratio $R$ is the key parameter to evalutate ability of the thermal switch, $R = \frac{{\kappa_{\rm on}}}{\kappa_{\rm off}}$,
where 
${\kappa_{\rm on}}$ is the largest $\kappa$ in the ``on'' state, and ${\kappa_{\rm off}}$ is the smallest $\kappa$ in the ``off'' state. %
%
%
Let us now attempt 
the the relationship between $R$ and strain at four different typical doping 
concentrations. %
For instance, at $5{\times}10^{-14} {\rm cm^{-2}}$ hole doping
concentration, $\kappa_{el}$ is equal to $8.97{\times}10^{-6}$ W/mK in
the equilibrium state, while 238.87 W/mK with only 1\% compressive strain. The total thermal conductivity 
changes from 8.92~W/mK without strain to 247.92~W/mK with 1\% compressive strain, 
achieving $R$ = 27.79, shown in Fig.~\ref{fig3}(b).
Furthermore, $R$ in compressive 
strain is much larger than that of tensile strain, indicating an obvious chirality of 
NPG. At a given strain, $R$ is directly proportional to the doping concentration, 
and the relative discrepancy of $R$ between different doping levels increases as a 
function of strain.

%
%
%

Referring to most of the former research about thermal management materials such as
${\rm VO_{2}}$, PE nanofiber, and other phase change 
materials~\cite{marinelli1998thermal, oh2010thermal, ho1972thermal, abdullaev1966thermal, zheng2019high, konstantinov2011isochoric, kim2016thermal, zheng2011reversible, crespi2014electrical, shin2019light, shrestha2019high}. To the best of our knowledge, we have collected $R$ of
all these interesting works combined with our NPG value, shown in Fig.~\ref{fig3}(c), %
indicating the potential of NPG for manufacturing high-performance thermal regulators. 

Remember that 
the mild change of ${\kappa}_{lat}$ can be originated from the unchanged space group of NPG with $P$6/$m$ (175)
during deformation, similar to VO$_2$ near the phase transition~\cite{berglund1969electronic}.
The physical reason for a wild change of ${\kappa}_{el}$ is the  
metal-insulator transition, since 
band structures and projected density of states (PDOS) change significantly shown in Fig.~\ref{fig4}(a-c). %

With deformation, %
conduction bands of NPG 
move downward while 
valence bands move upward, resulting in a smaller band gap compared with the 
equilibrium state with $E_g$ = 0.73~eV. %
Under 1\% compression, the band gap closes due to the decrease of distance 
between atoms, resulting in a shift of the electronic energy states. %
The projected band structure shows that the conduction band minimum is 
dominated by $p_z$ orbits of carbon and nitrogen atoms, while the valence 
band maximum is controlled by $p_x$ and $p_y$ orbitals of nitrogen 
atoms. Moreover, the band gap of NPG decreases slowly in expansion 
while declining rapidly in contraction. This phenomenon could be attributed to the 
chirality of NPG. We also find the main discrepancy of PDOS between the 
contraction state and the other two states is whether the overlap of 
$p_x$, $p_y$, and $p_z$ of nitrogen atoms exists, indicating the hybridization 
of orbits and delocalization of electrons. %
%
%

The band gap will fall to zero 
when $\Delta\theta$ reaches the critical angle of $-13.0^\circ$, indicating a metal-insulator transition 
occurs during deformation. The most adopted method to trigger it, in previous
reports, is applying a large 10\% mechanical deformation to adjust the distance between 
atoms and to redistribute electronic wavefunction, called Mott 
Transition~\cite{naumis2017electronic, ghorbani2013strain, chang2013orbital, scalise2012strain}. %
For NPG, the metal-insulator transition emerges  after only 1\% change of the lattice 
constant, which is much easier to be realized by substrate engineering, %
and could be used as super-sensitive 
pressure sensors 
in nanomaterials science. %

To explore electronic transport properties, we built NPG devices 
shown in Fig. S13. Without bias potential, transmission spectrums of different  
states are shown in Fig.~\ref{fig4}(d). 
Around the Fermi level, there is no 
transmission for all of them except the state of $\Delta\theta=-13.0^\circ$ and 
$\Delta\theta=-17.4^\circ$, corresponding semi-metal and metal states, respectively. %
Then electrical resistances of different states are invested by setting the 
bias voltage range from 0$\sim$1.0~V, shown in Fig.~\ref{fig4}(e). 
There are 7 orders of magnitude difference in 
electrical resistance between the state of $\Delta\theta=-17.4^\circ$ and the equilibrium 
state, while only 2 orders of magnitude difference between the state of 
$\Delta\theta=16.2^\circ$ and the equilibrium state. The details of calculation are shown in the Supporting Information~\cite{brandbyge2002density, smidstrup2019quantumatk, datta1996electronic}.


Let us now attempt the mechanism of the metal-insulator transition at the electronic 
distribution level. Fig.~\ref{fig4}(f) 
presents the charge density isosurface of NPG in three states. Contraction is more effective than an expansion to decrease the distance between adjacent nitrogen atoms, resulting in delocalizing the state 
and narrowing the band gap. %
This microscopic electronic behavior further supports the explanation of the metal-insulator transition in NPG. %
Except for the thermal and electrical properties, we also studied the optical property and negative thermal expansion~\cite{liu2019anisotropic, yoon2011negative} of NPG shown in Fig. S16 and Fig. S17. Furthermore, we validated the effective medium concept of NPG~\cite{liu2016continuum} and discussed the potential methods to obtain NPG~\cite{kambe2013pi, sakamoto2017coordination, treier2011surface, moreno2018bottom, sugimoto2015size, liu2018single, bieri2009porous, villalobos2020large, tan2016parent}.%

Before closing, we briefly discuss the energy cost to obtain such a thermal regulation.
The strain energy of NPG is presented in Fig. S18, and by changing the angle 
$\theta$ with the $\pm15^\circ$ from the equilibrium state, 
the largest energy needed is 140~meV/atom, %
about 3\% of the carbon bond-breaking energy~\cite{wang2015measurement}.  %
Thus, hexagonal rings in NPG are dynamically stable.
In \textit{ab initio} molecular dynamics calculation with two unit cells, the wrinkle along the z-axis is less than 0.05 nm in 300 K. For an extensive area of NPG, the vibration along the z-axis should be close to pure graphene, approximately equal to 1 nm ~\cite{meyer2007structure}, indicating the way of manipulating NPG is like graphene.
For instance, %
the strain is 0.0109 from the equilibrium state ($\theta=110.72^\circ$) to the 
expansion state ($\theta=120.27^\circ$), and the corresponding energy cost
per unit cell is 3.114~eV (the area of the unit cell is 2.2458~${\rm nm^{2}}$). 
The energy needed is 0.222~Jm$^{-2}$, even smaller than the cleavage 
energy of graphite with a value of 0.33~Jm$^{-2}$~\cite{wang2015measurement},
suggesting that NPG can be manipulated by a relatively small force. %

In summary, we have exhibited the demonstration of 
mapping from kirigami assembly on NPG metamaterials at atomically scale 
whose thermal switching ratio reaches 27.79 under only 1\% strain which is more 
than double the value of previous work.
It can be stemmed from a metal-insulator transition which switches the main heat carrier in NPG 
during the asymmetrical folding-unfolding deformation. %
We hope present results for inversely designing kirigami-inspired 
thermal switches do invigorate the studies aimed at uncovering intriguing 
high-performance thermal regulators, thermal diode, and thermal transistor, %
in the discipline of phononics, which will play essential
roles in nanoscale calorimeters and thermal management of 
microelectronics. %
\begin{acknowledgements}
We acknowledge the support from the National Natural Science Foundation of China 
(No. 12104356, 
52071258, 
52035008, 51825503) and the Tecent Foundation (XPLORER-2020-1035). 
Z.G. acknowledges the support of China Postdoctoral Science Foundation (No. 2022M712552) 
and the Fundamental Research Funds for the Central Universities. 
We also acknowledge the support by HPC Platform, Xi’an Jiaotong University. %
\end{acknowledgements}


 \bibliography{References} 

\begin{thebibliography}{89}%
\makeatletter
\providecommand \@ifxundefined [1]{%
 \@ifx{#1\undefined}
}%
\providecommand \@ifnum [1]{%
 \ifnum #1\expandafter \@firstoftwo
 \else \expandafter \@secondoftwo
 \fi
}%
\providecommand \@ifx [1]{%
 \ifx #1\expandafter \@firstoftwo
 \else \expandafter \@secondoftwo
 \fi
}%
\providecommand \natexlab [1]{#1}%
\providecommand \enquote  [1]{``#1''}%
\providecommand \bibnamefont  [1]{#1}%
\providecommand \bibfnamefont [1]{#1}%
\providecommand \citenamefont [1]{#1}%
\providecommand \href@noop [0]{\@secondoftwo}%
\providecommand \href [0]{\begingroup \@sanitize@url \@href}%
\providecommand \@href[1]{\@@startlink{#1}\@@href}%
\providecommand \@@href[1]{\endgroup#1\@@endlink}%
\providecommand \@sanitize@url [0]{\catcode `\\12\catcode `\$12\catcode
  `\&12\catcode `\#12\catcode `\^12\catcode `\_12\catcode `\%12\relax}%
\providecommand \@@startlink[1]{}%
\providecommand \@@endlink[0]{}%
\providecommand \url  [0]{\begingroup\@sanitize@url \@url }%
\providecommand \@url [1]{\endgroup\@href {#1}{\urlprefix }}%
\providecommand \urlprefix  [0]{URL }%
\providecommand \Eprint [0]{\href }%
\providecommand \doibase [0]{http://dx.doi.org/}%
\providecommand \selectlanguage [0]{\@gobble}%
\providecommand \bibinfo  [0]{\@secondoftwo}%
\providecommand \bibfield  [0]{\@secondoftwo}%
\providecommand \translation [1]{[#1]}%
\providecommand \BibitemOpen [0]{}%
\providecommand \bibitemStop [0]{}%
\providecommand \bibitemNoStop [0]{.\EOS\space}%
\providecommand \EOS [0]{\spacefactor3000\relax}%
\providecommand \BibitemShut  [1]{\csname bibitem#1\endcsname}%
\let\auto@bib@innerbib\@empty
\bibitem [{\citenamefont {Waldrop}(2016)}]{waldrop2016chips}%
  \BibitemOpen
  \bibfield  {author} {\bibinfo {author} {\bibfnamefont {M~Mitchell}\
  \bibnamefont {Waldrop}},\ }\bibfield  {title} {\bibinfo {title} {{The chips
  are down for Moore’s law}},\ }\href {\doibase 10.1038/530144a} {\bibfield
  {journal} {\bibinfo  {journal} {Nature}\ }\textbf {\bibinfo {volume} {530}},\
  \bibinfo {pages} {144--147} (\bibinfo {year} {2016})}\BibitemShut {NoStop}%
\bibitem [{\citenamefont {Ball}(2012)}]{Philip2012computer}%
  \BibitemOpen
  \bibfield  {author} {\bibinfo {author} {\bibfnamefont {Philip}\ \bibnamefont
  {Ball}},\ }\bibfield  {title} {\bibinfo {title} {{Computer engineering:
  Feeling the heat}},\ }\href {\doibase 10.1038/492174a} {\bibfield  {journal}
  {\bibinfo  {journal} {Nature}\ }\textbf {\bibinfo {volume} {492}},\ \bibinfo
  {pages} {174--176} (\bibinfo {year} {2012})}\BibitemShut {NoStop}%
\bibitem [{\citenamefont {Zhao}\ \emph {et~al.}(2014)\citenamefont {Zhao},
  \citenamefont {Lo}, \citenamefont {Zhang}, \citenamefont {Sun}, \citenamefont
  {Tan}, \citenamefont {Uher}, \citenamefont {Wolverton}, \citenamefont
  {Dravid},\ and\ \citenamefont {Kanatzidis}}]{zhao2014ultralow}%
  \BibitemOpen
  \bibfield  {author} {\bibinfo {author} {\bibfnamefont {Li-Dong}\ \bibnamefont
  {Zhao}}, \bibinfo {author} {\bibfnamefont {Shih-Han}\ \bibnamefont {Lo}},
  \bibinfo {author} {\bibfnamefont {Yongsheng}\ \bibnamefont {Zhang}}, \bibinfo
  {author} {\bibfnamefont {Hui}\ \bibnamefont {Sun}}, \bibinfo {author}
  {\bibfnamefont {Gangjian}\ \bibnamefont {Tan}}, \bibinfo {author}
  {\bibfnamefont {Ctirad}\ \bibnamefont {Uher}}, \bibinfo {author}
  {\bibfnamefont {Christopher}\ \bibnamefont {Wolverton}}, \bibinfo {author}
  {\bibfnamefont {Vinayak~P}\ \bibnamefont {Dravid}}, \ and\ \bibinfo {author}
  {\bibfnamefont {Mercouri~G}\ \bibnamefont {Kanatzidis}},\ }\bibfield  {title}
  {\bibinfo {title} {{Ultralow thermal conductivity and high thermoelectric
  figure of merit in SnSe crystals}},\ }\href {\doibase 10.1038/nature13184}
  {\bibfield  {journal} {\bibinfo  {journal} {Nature}\ }\textbf {\bibinfo
  {volume} {508}},\ \bibinfo {pages} {373--377} (\bibinfo {year}
  {2014})}\BibitemShut {NoStop}%
\bibitem [{\citenamefont {Jiang}\ \emph {et~al.}(2022)\citenamefont {Jiang},
  \citenamefont {Wang}, \citenamefont {Liu}, \citenamefont {Du}, \citenamefont
  {Li}, \citenamefont {Zhang}, \citenamefont {To}, \citenamefont {Wang},
  \citenamefont {Pan}, \citenamefont {Yu}, \citenamefont {Quéré},\ and\
  \citenamefont {Wang}}]{jiang2022inhibiting}%
  \BibitemOpen
  \bibfield  {author} {\bibinfo {author} {\bibfnamefont {Mengnan}\ \bibnamefont
  {Jiang}}, \bibinfo {author} {\bibfnamefont {Yang}\ \bibnamefont {Wang}},
  \bibinfo {author} {\bibfnamefont {Fayu}\ \bibnamefont {Liu}}, \bibinfo
  {author} {\bibfnamefont {Hanheng}\ \bibnamefont {Du}}, \bibinfo {author}
  {\bibfnamefont {Yuchao}\ \bibnamefont {Li}}, \bibinfo {author} {\bibfnamefont
  {Huanhuan}\ \bibnamefont {Zhang}}, \bibinfo {author} {\bibfnamefont {Suet}\
  \bibnamefont {To}}, \bibinfo {author} {\bibfnamefont {Steven}\ \bibnamefont
  {Wang}}, \bibinfo {author} {\bibfnamefont {Chin}\ \bibnamefont {Pan}},
  \bibinfo {author} {\bibfnamefont {Jihong}\ \bibnamefont {Yu}}, \bibinfo
  {author} {\bibfnamefont {David}\ \bibnamefont {Quéré}}, \ and\ \bibinfo
  {author} {\bibfnamefont {Zuankai}\ \bibnamefont {Wang}},\ }\bibfield  {title}
  {\bibinfo {title} {{Inhibiting the Leidenfrost effect above 1,000℃ for
  sustained thermal cooling}},\ }\href {\doibase 10.1038/s41586-021-04307-3}
  {\bibfield  {journal} {\bibinfo  {journal} {Nature}\ }\textbf {\bibinfo
  {volume} {601}},\ \bibinfo {pages} {568--572} (\bibinfo {year}
  {2022})}\BibitemShut {NoStop}%
\bibitem [{\citenamefont {Lindsay}\ \emph {et~al.}(2013)\citenamefont
  {Lindsay}, \citenamefont {Broido},\ and\ \citenamefont
  {Reinecke}}]{lindsay2013first}%
  \BibitemOpen
  \bibfield  {author} {\bibinfo {author} {\bibfnamefont {L.}~\bibnamefont
  {Lindsay}}, \bibinfo {author} {\bibfnamefont {D.~A.}\ \bibnamefont {Broido}},
  \ and\ \bibinfo {author} {\bibfnamefont {T.~L.}\ \bibnamefont {Reinecke}},\
  }\bibfield  {title} {\bibinfo {title} {{First-Principles Determination of
  Ultrahigh Thermal Conductivity of Boron Arsenide: A Competitor for
  Diamond?}}\ }\href {\doibase 10.1103/PhysRevLett.111.025901} {\bibfield
  {journal} {\bibinfo  {journal} {Phys. Rev. Lett.}\ }\textbf {\bibinfo
  {volume} {111}},\ \bibinfo {pages} {025901} (\bibinfo {year}
  {2013})}\BibitemShut {NoStop}%
\bibitem [{\citenamefont {Chen}\ \emph {et~al.}(2022)\citenamefont {Chen},
  \citenamefont {Xu}, \citenamefont {Zhou},\ and\ \citenamefont
  {Li}}]{chen2022interfacial}%
  \BibitemOpen
  \bibfield  {author} {\bibinfo {author} {\bibfnamefont {Jie}\ \bibnamefont
  {Chen}}, \bibinfo {author} {\bibfnamefont {Xiangfan}\ \bibnamefont {Xu}},
  \bibinfo {author} {\bibfnamefont {Jun}\ \bibnamefont {Zhou}}, \ and\ \bibinfo
  {author} {\bibfnamefont {Baowen}\ \bibnamefont {Li}},\ }\bibfield  {title}
  {\bibinfo {title} {{Interfacial thermal resistance: Past, present, and
  future}},\ }\href {\doibase 10.1103/RevModPhys.94.025002} {\bibfield
  {journal} {\bibinfo  {journal} {Rev. Mod. Phys.}\ }\textbf {\bibinfo {volume}
  {94}},\ \bibinfo {pages} {025002} (\bibinfo {year} {2022})}\BibitemShut
  {NoStop}%
\bibitem [{\citenamefont {Li}\ \emph {et~al.}(2012)\citenamefont {Li},
  \citenamefont {Ren}, \citenamefont {Wang}, \citenamefont {Zhang},
  \citenamefont {H\"anggi},\ and\ \citenamefont {Li}}]{li2012colloquium}%
  \BibitemOpen
  \bibfield  {author} {\bibinfo {author} {\bibfnamefont {Nianbei}\ \bibnamefont
  {Li}}, \bibinfo {author} {\bibfnamefont {Jie}\ \bibnamefont {Ren}}, \bibinfo
  {author} {\bibfnamefont {Lei}\ \bibnamefont {Wang}}, \bibinfo {author}
  {\bibfnamefont {Gang}\ \bibnamefont {Zhang}}, \bibinfo {author}
  {\bibfnamefont {Peter}\ \bibnamefont {H\"anggi}}, \ and\ \bibinfo {author}
  {\bibfnamefont {Baowen}\ \bibnamefont {Li}},\ }\bibfield  {title} {\bibinfo
  {title} {{Colloquium: Phononics: Manipulating heat flow with electronic
  analogs and beyond}},\ }\href {\doibase 10.1103/RevModPhys.84.1045}
  {\bibfield  {journal} {\bibinfo  {journal} {Rev. Mod. Phys.}\ }\textbf
  {\bibinfo {volume} {84}},\ \bibinfo {pages} {1045--1066} (\bibinfo {year}
  {2012})}\BibitemShut {NoStop}%
\bibitem [{\citenamefont {Gu}\ \emph {et~al.}(2018)\citenamefont {Gu},
  \citenamefont {Wei}, \citenamefont {Yin}, \citenamefont {Li},\ and\
  \citenamefont {Yang}}]{gu2018colloquium}%
  \BibitemOpen
  \bibfield  {author} {\bibinfo {author} {\bibfnamefont {Xiaokun}\ \bibnamefont
  {Gu}}, \bibinfo {author} {\bibfnamefont {Yujie}\ \bibnamefont {Wei}},
  \bibinfo {author} {\bibfnamefont {Xiaobo}\ \bibnamefont {Yin}}, \bibinfo
  {author} {\bibfnamefont {Baowen}\ \bibnamefont {Li}}, \ and\ \bibinfo
  {author} {\bibfnamefont {Ronggui}\ \bibnamefont {Yang}},\ }\bibfield  {title}
  {\bibinfo {title} {{Colloquium: Phononic thermal properties of
  two-dimensional materials}},\ }\href {\doibase 10.1103/RevModPhys.90.041002}
  {\bibfield  {journal} {\bibinfo  {journal} {Rev. Mod. Phys.}\ }\textbf
  {\bibinfo {volume} {90}},\ \bibinfo {pages} {041002} (\bibinfo {year}
  {2018})}\BibitemShut {NoStop}%
\bibitem [{\citenamefont {Li}\ \emph {et~al.}(2004)\citenamefont {Li},
  \citenamefont {Wang},\ and\ \citenamefont {Casati}}]{li2004thermal}%
  \BibitemOpen
  \bibfield  {author} {\bibinfo {author} {\bibfnamefont {Baowen}\ \bibnamefont
  {Li}}, \bibinfo {author} {\bibfnamefont {Lei}\ \bibnamefont {Wang}}, \ and\
  \bibinfo {author} {\bibfnamefont {Giulio}\ \bibnamefont {Casati}},\
  }\bibfield  {title} {\bibinfo {title} {{Thermal Diode: Rectification of Heat
  Flux}},\ }\href {\doibase 10.1103/PhysRevLett.93.184301} {\bibfield
  {journal} {\bibinfo  {journal} {Phys. Rev. Lett.}\ }\textbf {\bibinfo
  {volume} {93}},\ \bibinfo {pages} {184301} (\bibinfo {year}
  {2004})}\BibitemShut {NoStop}%
\bibitem [{\citenamefont {Wang}\ and\ \citenamefont
  {Li}(2007)}]{wang2007thermal}%
  \BibitemOpen
  \bibfield  {author} {\bibinfo {author} {\bibfnamefont {Lei}\ \bibnamefont
  {Wang}}\ and\ \bibinfo {author} {\bibfnamefont {Baowen}\ \bibnamefont {Li}},\
  }\bibfield  {title} {\bibinfo {title} {{Thermal Logic Gates: Computation with
  Phonons}},\ }\href {\doibase 10.1103/PhysRevLett.99.177208} {\bibfield
  {journal} {\bibinfo  {journal} {Phys. Rev. Lett.}\ }\textbf {\bibinfo
  {volume} {99}},\ \bibinfo {pages} {177208} (\bibinfo {year}
  {2007})}\BibitemShut {NoStop}%
\bibitem [{\citenamefont {Chang}\ \emph {et~al.}(2006)\citenamefont {Chang},
  \citenamefont {Okawa}, \citenamefont {Majumdar},\ and\ \citenamefont
  {Zettl}}]{chang2006solid}%
  \BibitemOpen
  \bibfield  {author} {\bibinfo {author} {\bibfnamefont {Chih~Wei}\
  \bibnamefont {Chang}}, \bibinfo {author} {\bibfnamefont {D}~\bibnamefont
  {Okawa}}, \bibinfo {author} {\bibfnamefont {A}~\bibnamefont {Majumdar}}, \
  and\ \bibinfo {author} {\bibfnamefont {A}~\bibnamefont {Zettl}},\ }\bibfield
  {title} {\bibinfo {title} {{Solid-state thermal rectifier}},\ }\href
  {\doibase 10.1126/science.1132898} {\bibfield  {journal} {\bibinfo  {journal}
  {Science}\ }\textbf {\bibinfo {volume} {314}},\ \bibinfo {pages} {1121--1124}
  (\bibinfo {year} {2006})}\BibitemShut {NoStop}%
\bibitem [{\citenamefont {Mart{\'\i}nez-P{\'e}rez}\ \emph
  {et~al.}(2015)\citenamefont {Mart{\'\i}nez-P{\'e}rez}, \citenamefont
  {Fornieri},\ and\ \citenamefont {Giazotto}}]{martinez2015rectification}%
  \BibitemOpen
  \bibfield  {author} {\bibinfo {author} {\bibfnamefont {Maria~Jos{\'e}}\
  \bibnamefont {Mart{\'\i}nez-P{\'e}rez}}, \bibinfo {author} {\bibfnamefont
  {Antonio}\ \bibnamefont {Fornieri}}, \ and\ \bibinfo {author} {\bibfnamefont
  {Francesco}\ \bibnamefont {Giazotto}},\ }\bibfield  {title} {\bibinfo {title}
  {{Rectification of electronic heat current by a hybrid thermal diode}},\
  }\href {https://doi.org/10.1038/nnano.2015.11} {\bibfield  {journal}
  {\bibinfo  {journal} {Nat. Nanotechnol.}\ }\textbf {\bibinfo {volume} {10}},\
  \bibinfo {pages} {303--307} (\bibinfo {year} {2015})}\BibitemShut {NoStop}%
\bibitem [{\citenamefont {Wang}\ and\ \citenamefont
  {Li}(2008)}]{wang2008thermal}%
  \BibitemOpen
  \bibfield  {author} {\bibinfo {author} {\bibfnamefont {Lei}\ \bibnamefont
  {Wang}}\ and\ \bibinfo {author} {\bibfnamefont {Baowen}\ \bibnamefont {Li}},\
  }\bibfield  {title} {\bibinfo {title} {{Thermal Memory: A Storage of Phononic
  Information}},\ }\href {\doibase 10.1103/PhysRevLett.101.267203} {\bibfield
  {journal} {\bibinfo  {journal} {Phys. Rev. Lett.}\ }\textbf {\bibinfo
  {volume} {101}},\ \bibinfo {pages} {267203} (\bibinfo {year}
  {2008})}\BibitemShut {NoStop}%
\bibitem [{\citenamefont {Wehmeyer}\ \emph {et~al.}(2017)\citenamefont
  {Wehmeyer}, \citenamefont {Yabuki}, \citenamefont {Monachon}, \citenamefont
  {Wu},\ and\ \citenamefont {Dames}}]{wehmeyer2017thermal}%
  \BibitemOpen
  \bibfield  {author} {\bibinfo {author} {\bibfnamefont {Geoff}\ \bibnamefont
  {Wehmeyer}}, \bibinfo {author} {\bibfnamefont {Tomohide}\ \bibnamefont
  {Yabuki}}, \bibinfo {author} {\bibfnamefont {Christian}\ \bibnamefont
  {Monachon}}, \bibinfo {author} {\bibfnamefont {Junqiao}\ \bibnamefont {Wu}},
  \ and\ \bibinfo {author} {\bibfnamefont {Chris}\ \bibnamefont {Dames}},\
  }\bibfield  {title} {\bibinfo {title} {{Thermal diodes, regulators, and
  switches: physical mechanisms and potential applications}},\ }\href
  {https://aip.scitation.org/doi/abs/10.1063/1.5001072} {\bibfield  {journal}
  {\bibinfo  {journal} {Appl. Phys. Rev.}\ }\textbf {\bibinfo {volume} {4}},\
  \bibinfo {pages} {041304} (\bibinfo {year} {2017})}\BibitemShut {NoStop}%
\bibitem [{\citenamefont {Li}\ \emph {et~al.}(2021)\citenamefont {Li},
  \citenamefont {Li}, \citenamefont {Han}, \citenamefont {Zheng}, \citenamefont
  {Li}, \citenamefont {Li}, \citenamefont {Fan},\ and\ \citenamefont
  {Qiu}}]{li2021transforming}%
  \BibitemOpen
  \bibfield  {author} {\bibinfo {author} {\bibfnamefont {Ying}\ \bibnamefont
  {Li}}, \bibinfo {author} {\bibfnamefont {Wei}\ \bibnamefont {Li}}, \bibinfo
  {author} {\bibfnamefont {Tiancheng}\ \bibnamefont {Han}}, \bibinfo {author}
  {\bibfnamefont {Xu}~\bibnamefont {Zheng}}, \bibinfo {author} {\bibfnamefont
  {Jiaxin}\ \bibnamefont {Li}}, \bibinfo {author} {\bibfnamefont {Baowen}\
  \bibnamefont {Li}}, \bibinfo {author} {\bibfnamefont {Shanhui}\ \bibnamefont
  {Fan}}, \ and\ \bibinfo {author} {\bibfnamefont {Cheng-Wei}\ \bibnamefont
  {Qiu}},\ }\bibfield  {title} {\bibinfo {title} {{Transforming heat transfer
  with thermal metamaterials and devices}},\ }\href
  {https://doi.org/10.1038/s41578-021-00283-2} {\bibfield  {journal} {\bibinfo
  {journal} {Nat. Rev. Mater.}\ }\textbf {\bibinfo {volume} {6}},\ \bibinfo
  {pages} {488--507} (\bibinfo {year} {2021})}\BibitemShut {NoStop}%
\bibitem [{\citenamefont {Jim{\'e}nez}\ \emph {et~al.}(2021)\citenamefont
  {Jim{\'e}nez}, \citenamefont {Groby},\ and\ \citenamefont
  {Romero-Garc{\'\i}a}}]{jimenez2021acoustic}%
  \BibitemOpen
  \bibfield  {author} {\bibinfo {author} {\bibfnamefont {No{\'e}}\ \bibnamefont
  {Jim{\'e}nez}}, \bibinfo {author} {\bibfnamefont {JP}~\bibnamefont {Groby}},
  \ and\ \bibinfo {author} {\bibfnamefont {V}~\bibnamefont
  {Romero-Garc{\'\i}a}},\ }\bibfield  {title} {\bibinfo {title} {{Acoustic
  waves in periodic structures, metamaterials, and porous media}},\ }\href
  {https://link.springer.com/book/10.1007/978-3-030-84300-7} {\bibfield
  {journal} {\bibinfo  {journal} {Ch. the Transfer Matrix Method in Acoustics,
  Springer International Publishing, Cham}\ ,\ \bibinfo {pages} {103--164}}
  (\bibinfo {year} {2021})}\BibitemShut {NoStop}%
\bibitem [{\citenamefont {Hao}\ \emph {et~al.}(2018)\citenamefont {Hao},
  \citenamefont {Li}, \citenamefont {Park}, \citenamefont {Moura},\ and\
  \citenamefont {Dames}}]{hao2018efficient}%
  \BibitemOpen
  \bibfield  {author} {\bibinfo {author} {\bibfnamefont {Menglong}\
  \bibnamefont {Hao}}, \bibinfo {author} {\bibfnamefont {Jian}\ \bibnamefont
  {Li}}, \bibinfo {author} {\bibfnamefont {Saehong}\ \bibnamefont {Park}},
  \bibinfo {author} {\bibfnamefont {Scott}\ \bibnamefont {Moura}}, \ and\
  \bibinfo {author} {\bibfnamefont {Chris}\ \bibnamefont {Dames}},\ }\bibfield
  {title} {\bibinfo {title} {{Efficient thermal management of Li-ion batteries
  with a passive interfacial thermal regulator based on a shape memory
  alloy}},\ }\href {https://www.nature.com/articles/s41560-018-0243-8}
  {\bibfield  {journal} {\bibinfo  {journal} {Nat. Energy}\ }\textbf {\bibinfo
  {volume} {3}},\ \bibinfo {pages} {899--906} (\bibinfo {year}
  {2018})}\BibitemShut {NoStop}%
\bibitem [{\citenamefont {Li}\ \emph {et~al.}(2015)\citenamefont {Li},
  \citenamefont {Shen}, \citenamefont {Wu}, \citenamefont {Huang},
  \citenamefont {Chen}, \citenamefont {Ni},\ and\ \citenamefont
  {Huang}}]{li2015temperature}%
  \BibitemOpen
  \bibfield  {author} {\bibinfo {author} {\bibfnamefont {Ying}\ \bibnamefont
  {Li}}, \bibinfo {author} {\bibfnamefont {Xiangying}\ \bibnamefont {Shen}},
  \bibinfo {author} {\bibfnamefont {Zuhui}\ \bibnamefont {Wu}}, \bibinfo
  {author} {\bibfnamefont {Junying}\ \bibnamefont {Huang}}, \bibinfo {author}
  {\bibfnamefont {Yixuan}\ \bibnamefont {Chen}}, \bibinfo {author}
  {\bibfnamefont {Yushan}\ \bibnamefont {Ni}}, \ and\ \bibinfo {author}
  {\bibfnamefont {Jiping}\ \bibnamefont {Huang}},\ }\bibfield  {title}
  {\bibinfo {title} {{Temperature-dependent transformation thermotics: from
  switchable thermal cloaks to macroscopic thermal diodes}},\ }\href
  {https://journals.aps.org/prl/abstract/10.1103/PhysRevLett.115.195503}
  {\bibfield  {journal} {\bibinfo  {journal} {Phys. Rev. Lett.}\ }\textbf
  {\bibinfo {volume} {115}},\ \bibinfo {pages} {195503} (\bibinfo {year}
  {2015})}\BibitemShut {NoStop}%
\bibitem [{\citenamefont {Bode}\ \emph {et~al.}(2004)\citenamefont {Bode},
  \citenamefont {Pietzsch}, \citenamefont {Kubetzka},\ and\ \citenamefont
  {Wiesendanger}}]{bode2004shape}%
  \BibitemOpen
  \bibfield  {author} {\bibinfo {author} {\bibfnamefont {M.}~\bibnamefont
  {Bode}}, \bibinfo {author} {\bibfnamefont {O.}~\bibnamefont {Pietzsch}},
  \bibinfo {author} {\bibfnamefont {A.}~\bibnamefont {Kubetzka}}, \ and\
  \bibinfo {author} {\bibfnamefont {R.}~\bibnamefont {Wiesendanger}},\
  }\bibfield  {title} {\bibinfo {title} {{Shape-Dependent Thermal Switching
  Behavior of Superparamagnetic Nanoislands}},\ }\href {\doibase
  10.1103/PhysRevLett.92.067201} {\bibfield  {journal} {\bibinfo  {journal}
  {Phys. Rev. Lett.}\ }\textbf {\bibinfo {volume} {92}},\ \bibinfo {pages}
  {067201} (\bibinfo {year} {2004})}\BibitemShut {NoStop}%
\bibitem [{\citenamefont {Zhang}\ \emph {et~al.}(2022)\citenamefont {Zhang},
  \citenamefont {Lv}, \citenamefont {Wang}, \citenamefont {Zhao}, \citenamefont
  {Xiong}, \citenamefont {Lv},\ and\ \citenamefont
  {Zhang}}]{zhang2022simultaneous}%
  \BibitemOpen
  \bibfield  {author} {\bibinfo {author} {\bibfnamefont {Yufeng}\ \bibnamefont
  {Zhang}}, \bibinfo {author} {\bibfnamefont {Qian}\ \bibnamefont {Lv}},
  \bibinfo {author} {\bibfnamefont {Haidong}\ \bibnamefont {Wang}}, \bibinfo
  {author} {\bibfnamefont {Shuaiyi}\ \bibnamefont {Zhao}}, \bibinfo {author}
  {\bibfnamefont {Qihua}\ \bibnamefont {Xiong}}, \bibinfo {author}
  {\bibfnamefont {Ruitao}\ \bibnamefont {Lv}}, \ and\ \bibinfo {author}
  {\bibfnamefont {Xing}\ \bibnamefont {Zhang}},\ }\bibfield  {title} {\bibinfo
  {title} {{Simultaneous electrical and thermal rectification in a monolayer
  lateral heterojunction}},\ }\href
  {https://www.science.org/doi/abs/10.1126/science.abq0883} {\bibfield
  {journal} {\bibinfo  {journal} {Science}\ }\textbf {\bibinfo {volume}
  {378}},\ \bibinfo {pages} {169--175} (\bibinfo {year} {2022})}\BibitemShut
  {NoStop}%
\bibitem [{\citenamefont {Thompson}\ \emph {et~al.}(2020)\citenamefont
  {Thompson}, \citenamefont {Zhu}, \citenamefont {Meyhofer},\ and\
  \citenamefont {Reddy}}]{thompson2020nanoscale}%
  \BibitemOpen
  \bibfield  {author} {\bibinfo {author} {\bibfnamefont {Dakotah}\ \bibnamefont
  {Thompson}}, \bibinfo {author} {\bibfnamefont {Linxiao}\ \bibnamefont {Zhu}},
  \bibinfo {author} {\bibfnamefont {Edgar}\ \bibnamefont {Meyhofer}}, \ and\
  \bibinfo {author} {\bibfnamefont {Pramod}\ \bibnamefont {Reddy}},\ }\bibfield
   {title} {\bibinfo {title} {{Nanoscale radiative thermal switching via
  multi-body effects}},\ }\href {https://doi.org/10.1038/s41565-019-0595-7}
  {\bibfield  {journal} {\bibinfo  {journal} {Nat. Nanotechnol.}\ }\textbf
  {\bibinfo {volume} {15}},\ \bibinfo {pages} {99--104} (\bibinfo {year}
  {2020})}\BibitemShut {NoStop}%
\bibitem [{\citenamefont {Langenberg}\ \emph {et~al.}(2019)\citenamefont
  {Langenberg}, \citenamefont {Saha}, \citenamefont {Holtz}, \citenamefont
  {Wang}, \citenamefont {Bugallo}, \citenamefont {Ferreiro-Vila}, \citenamefont
  {Paik}, \citenamefont {Hanke}, \citenamefont {Ganschow}, \citenamefont
  {Muller}, \citenamefont {Chen}, \citenamefont {Catalan}, \citenamefont
  {Domingo}, \citenamefont {Malen}, \citenamefont {Schlom},\ and\ \citenamefont
  {Rivadulla}}]{langenberg2019ferroelectric}%
  \BibitemOpen
  \bibfield  {author} {\bibinfo {author} {\bibfnamefont {Eric}\ \bibnamefont
  {Langenberg}}, \bibinfo {author} {\bibfnamefont {Dipanjan}\ \bibnamefont
  {Saha}}, \bibinfo {author} {\bibfnamefont {Megan~E.}\ \bibnamefont {Holtz}},
  \bibinfo {author} {\bibfnamefont {Jian-Jun}\ \bibnamefont {Wang}}, \bibinfo
  {author} {\bibfnamefont {David}\ \bibnamefont {Bugallo}}, \bibinfo {author}
  {\bibfnamefont {Elias}\ \bibnamefont {Ferreiro-Vila}}, \bibinfo {author}
  {\bibfnamefont {Hanjong}\ \bibnamefont {Paik}}, \bibinfo {author}
  {\bibfnamefont {Isabelle}\ \bibnamefont {Hanke}}, \bibinfo {author}
  {\bibfnamefont {Steffen}\ \bibnamefont {Ganschow}}, \bibinfo {author}
  {\bibfnamefont {David~A.}\ \bibnamefont {Muller}}, \bibinfo {author}
  {\bibfnamefont {Long-Qing}\ \bibnamefont {Chen}}, \bibinfo {author}
  {\bibfnamefont {Gustau}\ \bibnamefont {Catalan}}, \bibinfo {author}
  {\bibfnamefont {Neus}\ \bibnamefont {Domingo}}, \bibinfo {author}
  {\bibfnamefont {Jonathan}\ \bibnamefont {Malen}}, \bibinfo {author}
  {\bibfnamefont {Darrell~G.}\ \bibnamefont {Schlom}}, \ and\ \bibinfo {author}
  {\bibfnamefont {Francisco}\ \bibnamefont {Rivadulla}},\ }\bibfield  {title}
  {\bibinfo {title} {{Ferroelectric domain walls in PbTiO$_3$ are effective
  regulators of heat flow at room temperature}},\ }\href {\doibase
  10.1021/acs.nanolett.9b02991} {\bibfield  {journal} {\bibinfo  {journal}
  {Nano Lett.}\ }\textbf {\bibinfo {volume} {19}},\ \bibinfo {pages}
  {7901--7907} (\bibinfo {year} {2019})}\BibitemShut {NoStop}%
\bibitem [{\citenamefont {Du}\ \emph {et~al.}(2021)\citenamefont {Du},
  \citenamefont {Xiong}, \citenamefont {Delgado}, \citenamefont {Liao},
  \citenamefont {Peoples}, \citenamefont {Kantharaj}, \citenamefont
  {Chowdhury}, \citenamefont {Marconnet},\ and\ \citenamefont
  {Ruan}}]{du2021wide}%
  \BibitemOpen
  \bibfield  {author} {\bibinfo {author} {\bibfnamefont {Tingting}\
  \bibnamefont {Du}}, \bibinfo {author} {\bibfnamefont {Zixin}\ \bibnamefont
  {Xiong}}, \bibinfo {author} {\bibfnamefont {Luis}\ \bibnamefont {Delgado}},
  \bibinfo {author} {\bibfnamefont {Weizhi}\ \bibnamefont {Liao}}, \bibinfo
  {author} {\bibfnamefont {Joseph}\ \bibnamefont {Peoples}}, \bibinfo {author}
  {\bibfnamefont {Rajath}\ \bibnamefont {Kantharaj}}, \bibinfo {author}
  {\bibfnamefont {Prabudhya~Roy}\ \bibnamefont {Chowdhury}}, \bibinfo {author}
  {\bibfnamefont {Amy}\ \bibnamefont {Marconnet}}, \ and\ \bibinfo {author}
  {\bibfnamefont {Xiulin}\ \bibnamefont {Ruan}},\ }\bibfield  {title} {\bibinfo
  {title} {{Wide range continuously tunable and fast thermal switching based on
  compressible graphene composite foams}},\ }\href
  {https://doi.org/10.1038/s41467-021-25083-8} {\bibfield  {journal} {\bibinfo
  {journal} {Nat. Commun.}\ }\textbf {\bibinfo {volume} {12}},\ \bibinfo
  {pages} {1--10} (\bibinfo {year} {2021})}\BibitemShut {NoStop}%
\bibitem [{\citenamefont {Nonoyama}\ \emph {et~al.}(2020)\citenamefont
  {Nonoyama}, \citenamefont {Lee}, \citenamefont {Ota}, \citenamefont
  {Fujioka}, \citenamefont {Hong},\ and\ \citenamefont
  {Gong}}]{nonoyama2020instant}%
  \BibitemOpen
  \bibfield  {author} {\bibinfo {author} {\bibfnamefont {Takayuki}\
  \bibnamefont {Nonoyama}}, \bibinfo {author} {\bibfnamefont {Yong~Woo}\
  \bibnamefont {Lee}}, \bibinfo {author} {\bibfnamefont {Kumi}\ \bibnamefont
  {Ota}}, \bibinfo {author} {\bibfnamefont {Keigo}\ \bibnamefont {Fujioka}},
  \bibinfo {author} {\bibfnamefont {Wei}\ \bibnamefont {Hong}}, \ and\ \bibinfo
  {author} {\bibfnamefont {Jian~Ping}\ \bibnamefont {Gong}},\ }\bibfield
  {title} {\bibinfo {title} {{Instant thermal switching from soft hydrogel to
  rigid plastics inspired by thermophile proteins}},\ }\href
  {https://doi.org/10.1002/adma.201905878} {\bibfield  {journal} {\bibinfo
  {journal} {Adv. Mater.}\ }\textbf {\bibinfo {volume} {32}},\ \bibinfo {pages}
  {1905878} (\bibinfo {year} {2020})}\BibitemShut {NoStop}%
\bibitem [{\citenamefont {Ilic}\ \emph {et~al.}(2018)\citenamefont {Ilic},
  \citenamefont {Thomas}, \citenamefont {Christensen}, \citenamefont
  {Sherrott}, \citenamefont {Soljacic}, \citenamefont {Minnich}, \citenamefont
  {Miller},\ and\ \citenamefont {Atwater}}]{ilic2018active}%
  \BibitemOpen
  \bibfield  {author} {\bibinfo {author} {\bibfnamefont {Ognjen}\ \bibnamefont
  {Ilic}}, \bibinfo {author} {\bibfnamefont {Nathan~H}\ \bibnamefont {Thomas}},
  \bibinfo {author} {\bibfnamefont {Thomas}\ \bibnamefont {Christensen}},
  \bibinfo {author} {\bibfnamefont {Michelle~C}\ \bibnamefont {Sherrott}},
  \bibinfo {author} {\bibfnamefont {Marin}\ \bibnamefont {Soljacic}}, \bibinfo
  {author} {\bibfnamefont {Austin~J}\ \bibnamefont {Minnich}}, \bibinfo
  {author} {\bibfnamefont {Owen~D}\ \bibnamefont {Miller}}, \ and\ \bibinfo
  {author} {\bibfnamefont {Harry~A}\ \bibnamefont {Atwater}},\ }\bibfield
  {title} {\bibinfo {title} {{Active radiative thermal switching with graphene
  plasmon resonators}},\ }\href {https://doi.org/10.1021/acsnano.7b08231}
  {\bibfield  {journal} {\bibinfo  {journal} {ACS Nano}\ }\textbf {\bibinfo
  {volume} {12}},\ \bibinfo {pages} {2474--2481} (\bibinfo {year}
  {2018})}\BibitemShut {NoStop}%
\bibitem [{\citenamefont {Azulay}\ \emph {et~al.}(2003)\citenamefont {Azulay},
  \citenamefont {Eylon}, \citenamefont {Eshkenazi}, \citenamefont {Toker},
  \citenamefont {Balberg}, \citenamefont {Shimoni}, \citenamefont {Millo},\
  and\ \citenamefont {Balberg}}]{azulay2003electrical}%
  \BibitemOpen
  \bibfield  {author} {\bibinfo {author} {\bibfnamefont {D.}~\bibnamefont
  {Azulay}}, \bibinfo {author} {\bibfnamefont {M.}~\bibnamefont {Eylon}},
  \bibinfo {author} {\bibfnamefont {O.}~\bibnamefont {Eshkenazi}}, \bibinfo
  {author} {\bibfnamefont {D.}~\bibnamefont {Toker}}, \bibinfo {author}
  {\bibfnamefont {M.}~\bibnamefont {Balberg}}, \bibinfo {author} {\bibfnamefont
  {N.}~\bibnamefont {Shimoni}}, \bibinfo {author} {\bibfnamefont
  {O.}~\bibnamefont {Millo}}, \ and\ \bibinfo {author} {\bibfnamefont
  {I.}~\bibnamefont {Balberg}},\ }\bibfield  {title} {\bibinfo {title}
  {{Electrical-Thermal Switching in Carbon-Black--Polymer Composites as a Local
  Effect}},\ }\href {\doibase 10.1103/PhysRevLett.90.236601} {\bibfield
  {journal} {\bibinfo  {journal} {Phys. Rev. Lett.}\ }\textbf {\bibinfo
  {volume} {90}},\ \bibinfo {pages} {236601} (\bibinfo {year}
  {2003})}\BibitemShut {NoStop}%
\bibitem [{\citenamefont {Ihlefeld}\ \emph {et~al.}(2015)\citenamefont
  {Ihlefeld}, \citenamefont {Foley}, \citenamefont {Scrymgeour}, \citenamefont
  {Michael}, \citenamefont {McKenzie}, \citenamefont {Medlin}, \citenamefont
  {Wallace}, \citenamefont {Trolier-McKinstry},\ and\ \citenamefont
  {Hopkins}}]{ihlefeld2015room}%
  \BibitemOpen
  \bibfield  {author} {\bibinfo {author} {\bibfnamefont {Jon~F}\ \bibnamefont
  {Ihlefeld}}, \bibinfo {author} {\bibfnamefont {Brian~M}\ \bibnamefont
  {Foley}}, \bibinfo {author} {\bibfnamefont {David~A}\ \bibnamefont
  {Scrymgeour}}, \bibinfo {author} {\bibfnamefont {Joseph~R}\ \bibnamefont
  {Michael}}, \bibinfo {author} {\bibfnamefont {Bonnie~B}\ \bibnamefont
  {McKenzie}}, \bibinfo {author} {\bibfnamefont {Douglas~L}\ \bibnamefont
  {Medlin}}, \bibinfo {author} {\bibfnamefont {Margeaux}\ \bibnamefont
  {Wallace}}, \bibinfo {author} {\bibfnamefont {Susan}\ \bibnamefont
  {Trolier-McKinstry}}, \ and\ \bibinfo {author} {\bibfnamefont {Patrick~E}\
  \bibnamefont {Hopkins}},\ }\bibfield  {title} {\bibinfo {title}
  {{Room-temperature voltage tunable phonon thermal conductivity via
  reconfigurable interfaces in ferroelectric thin films}},\ }\href
  {https://doi.org/10.1021/nl504505t} {\bibfield  {journal} {\bibinfo
  {journal} {Nano Lett.}\ }\textbf {\bibinfo {volume} {15}},\ \bibinfo {pages}
  {1791--1795} (\bibinfo {year} {2015})}\BibitemShut {NoStop}%
\bibitem [{\citenamefont {Liu}\ \emph {et~al.}(2019{\natexlab{a}})\citenamefont
  {Liu}, \citenamefont {Chen},\ and\ \citenamefont {Dames}}]{liu2019electric}%
  \BibitemOpen
  \bibfield  {author} {\bibinfo {author} {\bibfnamefont {Chenhan}\ \bibnamefont
  {Liu}}, \bibinfo {author} {\bibfnamefont {Yunfei}\ \bibnamefont {Chen}}, \
  and\ \bibinfo {author} {\bibfnamefont {Chris}\ \bibnamefont {Dames}},\
  }\bibfield  {title} {\bibinfo {title} {{Electric-Field-Controlled Thermal
  Switch in Ferroelectric Materials Using First-Principles Calculations and
  Domain-Wall Engineering}},\ }\href {\doibase
  10.1103/PhysRevApplied.11.044002} {\bibfield  {journal} {\bibinfo  {journal}
  {Phys. Rev. Applied}\ }\textbf {\bibinfo {volume} {11}},\ \bibinfo {pages}
  {044002} (\bibinfo {year} {2019}{\natexlab{a}})}\BibitemShut {NoStop}%
\bibitem [{\citenamefont {Shin}\ \emph {et~al.}(2016)\citenamefont {Shin},
  \citenamefont {Kang}, \citenamefont {Tsai}, \citenamefont {Leal},
  \citenamefont {Braun},\ and\ \citenamefont {Cahill}}]{shin2016thermally}%
  \BibitemOpen
  \bibfield  {author} {\bibinfo {author} {\bibfnamefont {Jungwoo}\ \bibnamefont
  {Shin}}, \bibinfo {author} {\bibfnamefont {Minjee}\ \bibnamefont {Kang}},
  \bibinfo {author} {\bibfnamefont {Tsunghan}\ \bibnamefont {Tsai}}, \bibinfo
  {author} {\bibfnamefont {Cecilia}\ \bibnamefont {Leal}}, \bibinfo {author}
  {\bibfnamefont {Paul~V}\ \bibnamefont {Braun}}, \ and\ \bibinfo {author}
  {\bibfnamefont {David~G}\ \bibnamefont {Cahill}},\ }\bibfield  {title}
  {\bibinfo {title} {{Thermally functional liquid crystal networks by magnetic
  field driven molecular orientation}},\ }\href
  {https://doi.org/10.1021/acsmacrolett.6b00475} {\bibfield  {journal}
  {\bibinfo  {journal} {ACS Macro Lett.}\ }\textbf {\bibinfo {volume} {5}},\
  \bibinfo {pages} {955--960} (\bibinfo {year} {2016})}\BibitemShut {NoStop}%
\bibitem [{\citenamefont {Shrestha}\ \emph {et~al.}(2019)\citenamefont
  {Shrestha}, \citenamefont {Luan}, \citenamefont {Shin}, \citenamefont
  {Zhang}, \citenamefont {Luo}, \citenamefont {Lundh}, \citenamefont {Gong},
  \citenamefont {Bockstaller}, \citenamefont {Choi}, \citenamefont {Luo},
  \citenamefont {Chen}, \citenamefont {Hippalgaonkar},\ and\ \citenamefont
  {Shen}}]{shrestha2019high}%
  \BibitemOpen
  \bibfield  {author} {\bibinfo {author} {\bibfnamefont {Ramesh}\ \bibnamefont
  {Shrestha}}, \bibinfo {author} {\bibfnamefont {Yuxuan}\ \bibnamefont {Luan}},
  \bibinfo {author} {\bibfnamefont {Sunmi}\ \bibnamefont {Shin}}, \bibinfo
  {author} {\bibfnamefont {Teng}\ \bibnamefont {Zhang}}, \bibinfo {author}
  {\bibfnamefont {Xiao}\ \bibnamefont {Luo}}, \bibinfo {author} {\bibfnamefont
  {James~S.}\ \bibnamefont {Lundh}}, \bibinfo {author} {\bibfnamefont {Wei}\
  \bibnamefont {Gong}}, \bibinfo {author} {\bibfnamefont {Michael~R.}\
  \bibnamefont {Bockstaller}}, \bibinfo {author} {\bibfnamefont {Sukwon}\
  \bibnamefont {Choi}}, \bibinfo {author} {\bibfnamefont {Tengfei}\
  \bibnamefont {Luo}}, \bibinfo {author} {\bibfnamefont {Renkun}\ \bibnamefont
  {Chen}}, \bibinfo {author} {\bibfnamefont {Kedar}\ \bibnamefont
  {Hippalgaonkar}}, \ and\ \bibinfo {author} {\bibfnamefont {Sheng}\
  \bibnamefont {Shen}},\ }\bibfield  {title} {\bibinfo {title} {{High-contrast
  and reversible polymer thermal regulator by structural phase transition}},\
  }\href {https://www.science.org/doi/full/10.1126/sciadv.aax3777} {\bibfield
  {journal} {\bibinfo  {journal} {Sci. Adv.}\ }\textbf {\bibinfo {volume}
  {5}},\ \bibinfo {pages} {eaax3777} (\bibinfo {year} {2019})}\BibitemShut
  {NoStop}%
\bibitem [{\citenamefont {Aryana}\ \emph {et~al.}(2021)\citenamefont {Aryana},
  \citenamefont {Zhang}, \citenamefont {Tomko}, \citenamefont {Hoque},
  \citenamefont {Hoglund}, \citenamefont {Olson}, \citenamefont {Nag},
  \citenamefont {Read}, \citenamefont {R{\'\i}os}, \citenamefont {Hu},\ and\
  \citenamefont {Hopkins}}]{aryana2021suppressed}%
  \BibitemOpen
  \bibfield  {author} {\bibinfo {author} {\bibfnamefont {Kiumars}\ \bibnamefont
  {Aryana}}, \bibinfo {author} {\bibfnamefont {Yifei}\ \bibnamefont {Zhang}},
  \bibinfo {author} {\bibfnamefont {John~A}\ \bibnamefont {Tomko}}, \bibinfo
  {author} {\bibfnamefont {Md~Shafkat~Bin}\ \bibnamefont {Hoque}}, \bibinfo
  {author} {\bibfnamefont {Eric~R}\ \bibnamefont {Hoglund}}, \bibinfo {author}
  {\bibfnamefont {David~H}\ \bibnamefont {Olson}}, \bibinfo {author}
  {\bibfnamefont {Joyeeta}\ \bibnamefont {Nag}}, \bibinfo {author}
  {\bibfnamefont {John~C}\ \bibnamefont {Read}}, \bibinfo {author}
  {\bibfnamefont {Carlos}\ \bibnamefont {R{\'\i}os}}, \bibinfo {author}
  {\bibfnamefont {Juejun}\ \bibnamefont {Hu}}, \ and\ \bibinfo {author}
  {\bibfnamefont {Patrick~E.}\ \bibnamefont {Hopkins}},\ }\bibfield  {title}
  {\bibinfo {title} {{Suppressed electronic contribution in thermal
  conductivity of Ge$_2$Sb$_2$Se$_4$Te}},\ }\href
  {https://doi.org/10.1038/s41467-021-27121-x} {\bibfield  {journal} {\bibinfo
  {journal} {Nat. Commun.}\ }\textbf {\bibinfo {volume} {12}},\ \bibinfo
  {pages} {1--9} (\bibinfo {year} {2021})}\BibitemShut {NoStop}%
\bibitem [{\citenamefont {Cho}\ \emph {et~al.}(2014)\citenamefont {Cho},
  \citenamefont {Losego}, \citenamefont {Zhang}, \citenamefont {Kim},
  \citenamefont {Zuo}, \citenamefont {Petrov}, \citenamefont {Cahill},\ and\
  \citenamefont {Braun}}]{cho2014electrochemically}%
  \BibitemOpen
  \bibfield  {author} {\bibinfo {author} {\bibfnamefont {Jiung}\ \bibnamefont
  {Cho}}, \bibinfo {author} {\bibfnamefont {Mark~D}\ \bibnamefont {Losego}},
  \bibinfo {author} {\bibfnamefont {Hui~Gang}\ \bibnamefont {Zhang}}, \bibinfo
  {author} {\bibfnamefont {Honggyu}\ \bibnamefont {Kim}}, \bibinfo {author}
  {\bibfnamefont {Jianmin}\ \bibnamefont {Zuo}}, \bibinfo {author}
  {\bibfnamefont {Ivan}\ \bibnamefont {Petrov}}, \bibinfo {author}
  {\bibfnamefont {David~G}\ \bibnamefont {Cahill}}, \ and\ \bibinfo {author}
  {\bibfnamefont {Paul~V}\ \bibnamefont {Braun}},\ }\bibfield  {title}
  {\bibinfo {title} {{Electrochemically tunable thermal conductivity of lithium
  cobalt oxide}},\ }\href {https://doi.org/10.1038/ncomms5035} {\bibfield
  {journal} {\bibinfo  {journal} {Nat. Commun.}\ }\textbf {\bibinfo {volume}
  {5}},\ \bibinfo {pages} {1--6} (\bibinfo {year} {2014})}\BibitemShut
  {NoStop}%
\bibitem [{\citenamefont {Lu}\ \emph {et~al.}(2020)\citenamefont {Lu},
  \citenamefont {Huberman}, \citenamefont {Zhang}, \citenamefont {Song},
  \citenamefont {Wang}, \citenamefont {Vardar}, \citenamefont {Hunt},
  \citenamefont {Waluyo}, \citenamefont {Chen},\ and\ \citenamefont
  {Yildiz}}]{lu2020bi}%
  \BibitemOpen
  \bibfield  {author} {\bibinfo {author} {\bibfnamefont {Qiyang}\ \bibnamefont
  {Lu}}, \bibinfo {author} {\bibfnamefont {Samuel}\ \bibnamefont {Huberman}},
  \bibinfo {author} {\bibfnamefont {Hantao}\ \bibnamefont {Zhang}}, \bibinfo
  {author} {\bibfnamefont {Qichen}\ \bibnamefont {Song}}, \bibinfo {author}
  {\bibfnamefont {Jiayue}\ \bibnamefont {Wang}}, \bibinfo {author}
  {\bibfnamefont {Gulin}\ \bibnamefont {Vardar}}, \bibinfo {author}
  {\bibfnamefont {Adrian}\ \bibnamefont {Hunt}}, \bibinfo {author}
  {\bibfnamefont {Iradwikanari}\ \bibnamefont {Waluyo}}, \bibinfo {author}
  {\bibfnamefont {Gang}\ \bibnamefont {Chen}}, \ and\ \bibinfo {author}
  {\bibfnamefont {Bilge}\ \bibnamefont {Yildiz}},\ }\bibfield  {title}
  {\bibinfo {title} {{Bi-directional tuning of thermal transport in SrCoOx with
  electrochemically induced phase transitions}},\ }\href
  {https://doi.org/10.1038/s41563-020-0612-0} {\bibfield  {journal} {\bibinfo
  {journal} {Nat. Mater.}\ }\textbf {\bibinfo {volume} {19}},\ \bibinfo {pages}
  {655--662} (\bibinfo {year} {2020})}\BibitemShut {NoStop}%
\bibitem [{\citenamefont {Shin}\ \emph {et~al.}(2019)\citenamefont {Shin},
  \citenamefont {Sung}, \citenamefont {Kang}, \citenamefont {Xie},
  \citenamefont {Lee}, \citenamefont {Lee}, \citenamefont {White},
  \citenamefont {Leal}, \citenamefont {Sottos}, \citenamefont {Braun},\ and\
  \citenamefont {Cahill}}]{shin2019light}%
  \BibitemOpen
  \bibfield  {author} {\bibinfo {author} {\bibfnamefont {Jungwoo}\ \bibnamefont
  {Shin}}, \bibinfo {author} {\bibfnamefont {Jaeuk}\ \bibnamefont {Sung}},
  \bibinfo {author} {\bibfnamefont {Minjee}\ \bibnamefont {Kang}}, \bibinfo
  {author} {\bibfnamefont {Xu}~\bibnamefont {Xie}}, \bibinfo {author}
  {\bibfnamefont {Byeongdu}\ \bibnamefont {Lee}}, \bibinfo {author}
  {\bibfnamefont {Kyung~Min}\ \bibnamefont {Lee}}, \bibinfo {author}
  {\bibfnamefont {Timothy~J.}\ \bibnamefont {White}}, \bibinfo {author}
  {\bibfnamefont {Cecilia}\ \bibnamefont {Leal}}, \bibinfo {author}
  {\bibfnamefont {Nancy~R.}\ \bibnamefont {Sottos}}, \bibinfo {author}
  {\bibfnamefont {Paul~V.}\ \bibnamefont {Braun}}, \ and\ \bibinfo {author}
  {\bibfnamefont {David~G.}\ \bibnamefont {Cahill}},\ }\bibfield  {title}
  {\bibinfo {title} {{Light-triggered thermal conductivity switching in
  azobenzene polymers}},\ }\href
  {https://www.pnas.org/doi/abs/10.1073/pnas.1817082116} {\bibfield  {journal}
  {\bibinfo  {journal} {Proc. Natl Acad. Sci. USA}\ }\textbf {\bibinfo {volume}
  {116}},\ \bibinfo {pages} {5973--5978} (\bibinfo {year} {2019})}\BibitemShut
  {NoStop}%
\bibitem [{\citenamefont {Tomko}\ \emph {et~al.}(2018)\citenamefont {Tomko},
  \citenamefont {Pena-Francesch}, \citenamefont {Jung}, \citenamefont {Tyagi},
  \citenamefont {Allen}, \citenamefont {Demirel},\ and\ \citenamefont
  {Hopkins}}]{tomko2018tunable}%
  \BibitemOpen
  \bibfield  {author} {\bibinfo {author} {\bibfnamefont {John~A}\ \bibnamefont
  {Tomko}}, \bibinfo {author} {\bibfnamefont {Abdon}\ \bibnamefont
  {Pena-Francesch}}, \bibinfo {author} {\bibfnamefont {Huihun}\ \bibnamefont
  {Jung}}, \bibinfo {author} {\bibfnamefont {Madhusudan}\ \bibnamefont
  {Tyagi}}, \bibinfo {author} {\bibfnamefont {Benjamin~D}\ \bibnamefont
  {Allen}}, \bibinfo {author} {\bibfnamefont {Melik~C}\ \bibnamefont
  {Demirel}}, \ and\ \bibinfo {author} {\bibfnamefont {Patrick~E}\ \bibnamefont
  {Hopkins}},\ }\bibfield  {title} {\bibinfo {title} {{Tunable thermal
  transport and reversible thermal conductivity switching in topologically
  networked bio-inspired materials}},\ }\href
  {https://doi.org/10.1038/s41565-018-0227-7} {\bibfield  {journal} {\bibinfo
  {journal} {Nat. Nanotechnol.}\ }\textbf {\bibinfo {volume} {13}},\ \bibinfo
  {pages} {959--964} (\bibinfo {year} {2018})}\BibitemShut {NoStop}%
\bibitem [{\citenamefont {Efetov}\ and\ \citenamefont
  {Kim}(2010)}]{efetov2010controlling}%
  \BibitemOpen
  \bibfield  {author} {\bibinfo {author} {\bibfnamefont {Dmitri~K.}\
  \bibnamefont {Efetov}}\ and\ \bibinfo {author} {\bibfnamefont {Philip}\
  \bibnamefont {Kim}},\ }\bibfield  {title} {\bibinfo {title} {{Controlling
  Electron-Phonon Interactions in Graphene at Ultrahigh Carrier Densities}},\
  }\href {\doibase 10.1103/PhysRevLett.105.256805} {\bibfield  {journal}
  {\bibinfo  {journal} {Phys. Rev. Lett.}\ }\textbf {\bibinfo {volume} {105}},\
  \bibinfo {pages} {256805} (\bibinfo {year} {2010})}\BibitemShut {NoStop}%
\bibitem [{\citenamefont {Chuang}\ \emph {et~al.}(2014)\citenamefont {Chuang},
  \citenamefont {Tan}, \citenamefont {Ghimire}, \citenamefont {Perera},
  \citenamefont {Chamlagain}, \citenamefont {Cheng}, \citenamefont {Yan},
  \citenamefont {Mandrus}, \citenamefont {Tománek},\ and\ \citenamefont
  {Zhou}}]{chuang2014high}%
  \BibitemOpen
  \bibfield  {author} {\bibinfo {author} {\bibfnamefont {Hsun-Jen}\
  \bibnamefont {Chuang}}, \bibinfo {author} {\bibfnamefont {Xuebin}\
  \bibnamefont {Tan}}, \bibinfo {author} {\bibfnamefont {Nirmal~Jeevi}\
  \bibnamefont {Ghimire}}, \bibinfo {author} {\bibfnamefont
  {Meeghage~Madusanka}\ \bibnamefont {Perera}}, \bibinfo {author}
  {\bibfnamefont {Bhim}\ \bibnamefont {Chamlagain}}, \bibinfo {author}
  {\bibfnamefont {Mark Ming-Cheng}\ \bibnamefont {Cheng}}, \bibinfo {author}
  {\bibfnamefont {Jiaqiang}\ \bibnamefont {Yan}}, \bibinfo {author}
  {\bibfnamefont {David}\ \bibnamefont {Mandrus}}, \bibinfo {author}
  {\bibfnamefont {David}\ \bibnamefont {Tománek}}, \ and\ \bibinfo {author}
  {\bibfnamefont {Zhixian}\ \bibnamefont {Zhou}},\ }\bibfield  {title}
  {\bibinfo {title} {{High Mobility WSe$_2$ p- and n-Type Field-Effect
  Transistors Contacted by Highly Doped Graphene for Low-Resistance
  Contacts}},\ }\href {https://doi.org/10.1021/nl501275p} {\bibfield  {journal}
  {\bibinfo  {journal} {Nano Lett.}\ }\textbf {\bibinfo {volume} {14}},\
  \bibinfo {pages} {3594--3601} (\bibinfo {year} {2014})}\BibitemShut {NoStop}%
\bibitem [{\citenamefont {Marinelli}\ \emph {et~al.}(1998)\citenamefont
  {Marinelli}, \citenamefont {Mercuri}, \citenamefont {Zammit},\ and\
  \citenamefont {Scudieri}}]{marinelli1998thermal}%
  \BibitemOpen
  \bibfield  {author} {\bibinfo {author} {\bibfnamefont {M.}~\bibnamefont
  {Marinelli}}, \bibinfo {author} {\bibfnamefont {F.}~\bibnamefont {Mercuri}},
  \bibinfo {author} {\bibfnamefont {U.}~\bibnamefont {Zammit}}, \ and\ \bibinfo
  {author} {\bibfnamefont {F.}~\bibnamefont {Scudieri}},\ }\bibfield  {title}
  {\bibinfo {title} {{Thermal conductivity and thermal diffusivity of the
  cyanobiphenyl $(n\mathrm{CB})$ homologous series}},\ }\href {\doibase
  10.1103/PhysRevE.58.5860} {\bibfield  {journal} {\bibinfo  {journal} {Phys.
  Rev. E}\ }\textbf {\bibinfo {volume} {58}},\ \bibinfo {pages} {5860--5866}
  (\bibinfo {year} {1998})}\BibitemShut {NoStop}%
\bibitem [{\citenamefont {Oh}\ \emph {et~al.}(2010)\citenamefont {Oh},
  \citenamefont {Ko}, \citenamefont {Ramanathan},\ and\ \citenamefont
  {Cahill}}]{oh2010thermal}%
  \BibitemOpen
  \bibfield  {author} {\bibinfo {author} {\bibfnamefont {Dong-Wook}\
  \bibnamefont {Oh}}, \bibinfo {author} {\bibfnamefont {Changhyun}\
  \bibnamefont {Ko}}, \bibinfo {author} {\bibfnamefont {Shriram}\ \bibnamefont
  {Ramanathan}}, \ and\ \bibinfo {author} {\bibfnamefont {David~G}\
  \bibnamefont {Cahill}},\ }\bibfield  {title} {\bibinfo {title} {{Thermal
  conductivity and dynamic heat capacity across the metal-insulator transition
  in thin film VO$_2$}},\ }\href
  {https://aip.scitation.org/doi/abs/10.1063/1.3394016} {\bibfield  {journal}
  {\bibinfo  {journal} {Appl. Phys. Lett.}\ }\textbf {\bibinfo {volume} {96}},\
  \bibinfo {pages} {151906} (\bibinfo {year} {2010})}\BibitemShut {NoStop}%
\bibitem [{\citenamefont {Ho}\ \emph {et~al.}(1972)\citenamefont {Ho},
  \citenamefont {Powell},\ and\ \citenamefont {Liley}}]{ho1972thermal}%
  \BibitemOpen
  \bibfield  {author} {\bibinfo {author} {\bibfnamefont {Cho~Yen}\ \bibnamefont
  {Ho}}, \bibinfo {author} {\bibfnamefont {Reginald~W}\ \bibnamefont {Powell}},
  \ and\ \bibinfo {author} {\bibfnamefont {Peter~E}\ \bibnamefont {Liley}},\
  }\bibfield  {title} {\bibinfo {title} {{Thermal conductivity of the
  elements}},\ }\href {https://aip.scitation.org/doi/abs/10.1063/1.3253100}
  {\bibfield  {journal} {\bibinfo  {journal} {J. Phys. Chem. Ref. Data}\
  }\textbf {\bibinfo {volume} {1}},\ \bibinfo {pages} {279--421} (\bibinfo
  {year} {1972})}\BibitemShut {NoStop}%
\bibitem [{\citenamefont {Abdullaev}\ \emph {et~al.}(1966)\citenamefont
  {Abdullaev}, \citenamefont {Mekhtieva}, \citenamefont {Sh.~Abdinov},
  \citenamefont {Aliev},\ and\ \citenamefont {Alieva}}]{abdullaev1966thermal}%
  \BibitemOpen
  \bibfield  {author} {\bibinfo {author} {\bibfnamefont {GB}~\bibnamefont
  {Abdullaev}}, \bibinfo {author} {\bibfnamefont {SI}~\bibnamefont
  {Mekhtieva}}, \bibinfo {author} {\bibfnamefont {D}~\bibnamefont
  {Sh.~Abdinov}}, \bibinfo {author} {\bibfnamefont {GM}~\bibnamefont {Aliev}},
  \ and\ \bibinfo {author} {\bibfnamefont {SG}~\bibnamefont {Alieva}},\
  }\bibfield  {title} {\bibinfo {title} {{Thermal conductivity of Selenium}},\
  }\href {https://onlinelibrary.wiley.com/doi/abs/10.1002/pssb.19660130203}
  {\bibfield  {journal} {\bibinfo  {journal} {Phys. Status Solidi B}\ }\textbf
  {\bibinfo {volume} {13}},\ \bibinfo {pages} {315--323} (\bibinfo {year}
  {1966})}\BibitemShut {NoStop}%
\bibitem [{\citenamefont {Zheng}\ \emph {et~al.}(2019)\citenamefont {Zheng},
  \citenamefont {Zhu}, \citenamefont {Diao}, \citenamefont {Banerjee},\ and\
  \citenamefont {Cahill}}]{zheng2019high}%
  \BibitemOpen
  \bibfield  {author} {\bibinfo {author} {\bibfnamefont {Qiye}\ \bibnamefont
  {Zheng}}, \bibinfo {author} {\bibfnamefont {Gaohua}\ \bibnamefont {Zhu}},
  \bibinfo {author} {\bibfnamefont {Zhu}\ \bibnamefont {Diao}}, \bibinfo
  {author} {\bibfnamefont {Debasish}\ \bibnamefont {Banerjee}}, \ and\ \bibinfo
  {author} {\bibfnamefont {David~G}\ \bibnamefont {Cahill}},\ }\bibfield
  {title} {\bibinfo {title} {{High contrast thermal conductivity change in
  Ni--Mn--In heusler alloys near room temperature}},\ }\href
  {https://onlinelibrary.wiley.com/doi/abs/10.1002/adem.201801342} {\bibfield
  {journal} {\bibinfo  {journal} {Adv. Eng. Mater.}\ }\textbf {\bibinfo
  {volume} {21}},\ \bibinfo {pages} {1801342} (\bibinfo {year}
  {2019})}\BibitemShut {NoStop}%
\bibitem [{\citenamefont {Konstantinov}\ \emph {et~al.}(2011)\citenamefont
  {Konstantinov}, \citenamefont {Revyakin},\ and\ \citenamefont
  {Sagan}}]{konstantinov2011isochoric}%
  \BibitemOpen
  \bibfield  {author} {\bibinfo {author} {\bibfnamefont {VA}~\bibnamefont
  {Konstantinov}}, \bibinfo {author} {\bibfnamefont {VP}~\bibnamefont
  {Revyakin}}, \ and\ \bibinfo {author} {\bibfnamefont {VV}~\bibnamefont
  {Sagan}},\ }\bibfield  {title} {\bibinfo {title} {{Isochoric thermal
  conductivity of solid n-alkanes: hexane C$_6$H$_{14}$}},\ }\href
  {https://aip.scitation.org/doi/abs/10.1063/1.3604519} {\bibfield  {journal}
  {\bibinfo  {journal} {Low Temp. Phys.}\ }\textbf {\bibinfo {volume} {37}},\
  \bibinfo {pages} {420--423} (\bibinfo {year} {2011})}\BibitemShut {NoStop}%
\bibitem [{\citenamefont {Kim}\ and\ \citenamefont
  {Kaviany}(2016)}]{kim2016thermal}%
  \BibitemOpen
  \bibfield  {author} {\bibinfo {author} {\bibfnamefont {Kwangnam}\
  \bibnamefont {Kim}}\ and\ \bibinfo {author} {\bibfnamefont {Massoud}\
  \bibnamefont {Kaviany}},\ }\bibfield  {title} {\bibinfo {title} {{Thermal
  conductivity switch: Optimal semiconductor/metal melting transition}},\
  }\href {https://journals.aps.org/prb/abstract/10.1103/PhysRevB.94.155203}
  {\bibfield  {journal} {\bibinfo  {journal} {Phys. Rev. B}\ }\textbf {\bibinfo
  {volume} {94}},\ \bibinfo {pages} {155203} (\bibinfo {year}
  {2016})}\BibitemShut {NoStop}%
\bibitem [{\citenamefont {Zheng}\ \emph {et~al.}(2011)\citenamefont {Zheng},
  \citenamefont {Gao}, \citenamefont {Wang},\ and\ \citenamefont
  {Chen}}]{zheng2011reversible}%
  \BibitemOpen
  \bibfield  {author} {\bibinfo {author} {\bibfnamefont {Ruiting}\ \bibnamefont
  {Zheng}}, \bibinfo {author} {\bibfnamefont {Jinwei}\ \bibnamefont {Gao}},
  \bibinfo {author} {\bibfnamefont {Jianjian}\ \bibnamefont {Wang}}, \ and\
  \bibinfo {author} {\bibfnamefont {Gang}\ \bibnamefont {Chen}},\ }\bibfield
  {title} {\bibinfo {title} {{Reversible temperature regulation of electrical
  and thermal conductivity using liquid--solid phase transitions}},\ }\href
  {https://www.nature.com/articles/ncomms1288} {\bibfield  {journal} {\bibinfo
  {journal} {Nat. Commun.}\ }\textbf {\bibinfo {volume} {2}},\ \bibinfo {pages}
  {289} (\bibinfo {year} {2011})}\BibitemShut {NoStop}%
\bibitem [{\citenamefont {Crespi}\ \emph {et~al.}(2014)\citenamefont {Crespi},
  \citenamefont {Ghetti}, \citenamefont {Boniardi},\ and\ \citenamefont
  {Lacaita}}]{crespi2014electrical}%
  \BibitemOpen
  \bibfield  {author} {\bibinfo {author} {\bibfnamefont {Luca}\ \bibnamefont
  {Crespi}}, \bibinfo {author} {\bibfnamefont {Andrea}\ \bibnamefont {Ghetti}},
  \bibinfo {author} {\bibfnamefont {Mattia}\ \bibnamefont {Boniardi}}, \ and\
  \bibinfo {author} {\bibfnamefont {Andrea~L}\ \bibnamefont {Lacaita}},\
  }\bibfield  {title} {\bibinfo {title} {{Electrical conductivity discontinuity
  at melt in phase change memory}},\ }\href
  {https://ieeexplore.ieee.org/abstract/document/6812189/} {\bibfield
  {journal} {\bibinfo  {journal} {IEEE Electron Device Lett.}\ }\textbf
  {\bibinfo {volume} {35}},\ \bibinfo {pages} {747--749} (\bibinfo {year}
  {2014})}\BibitemShut {NoStop}%
\bibitem [{\citenamefont {Baker}(1986)}]{baker1986limiting}%
  \BibitemOpen
  \bibfield  {author} {\bibinfo {author} {\bibfnamefont {J~Eddie}\ \bibnamefont
  {Baker}},\ }\bibfield  {title} {\bibinfo {title} {{Limiting positions of a
  Bricard linkage and their possible relevance to the cyclohexane molecule}},\
  }\href
  {https://www.sciencedirect.com/science/article/abs/pii/0094114X86901011}
  {\bibfield  {journal} {\bibinfo  {journal} {Mech. Mach. Theory}\ }\textbf
  {\bibinfo {volume} {21}},\ \bibinfo {pages} {253--260} (\bibinfo {year}
  {1986})}\BibitemShut {NoStop}%
\bibitem [{\citenamefont {Tsetseris}\ \emph {et~al.}(2014)\citenamefont
  {Tsetseris}, \citenamefont {Wang},\ and\ \citenamefont
  {Pantelides}}]{tsetseris2014substitutional}%
  \BibitemOpen
  \bibfield  {author} {\bibinfo {author} {\bibfnamefont {L.}~\bibnamefont
  {Tsetseris}}, \bibinfo {author} {\bibfnamefont {B.}~\bibnamefont {Wang}}, \
  and\ \bibinfo {author} {\bibfnamefont {S.~T.}\ \bibnamefont {Pantelides}},\
  }\bibfield  {title} {\bibinfo {title} {{Substitutional doping of graphene:
  The role of carbon divacancies}},\ }\href {\doibase
  10.1103/PhysRevB.89.035411} {\bibfield  {journal} {\bibinfo  {journal} {Phys.
  Rev. B}\ }\textbf {\bibinfo {volume} {89}},\ \bibinfo {pages} {035411}
  (\bibinfo {year} {2014})}\BibitemShut {NoStop}%
\bibitem [{\citenamefont {Bl\"ochl}(1994)}]{blochl1994projector}%
  \BibitemOpen
  \bibfield  {author} {\bibinfo {author} {\bibfnamefont {P.~E.}\ \bibnamefont
  {Bl\"ochl}},\ }\bibfield  {title} {\bibinfo {title} {{Projector
  augmented-wave method}},\ }\href {\doibase 10.1103/PhysRevB.50.17953}
  {\bibfield  {journal} {\bibinfo  {journal} {Phys. Rev. B}\ }\textbf {\bibinfo
  {volume} {50}},\ \bibinfo {pages} {17953--17979} (\bibinfo {year}
  {1994})}\BibitemShut {NoStop}%
\bibitem [{\citenamefont {Kresse}\ and\ \citenamefont
  {Joubert}(1999)}]{kresse1999ultrasoft}%
  \BibitemOpen
  \bibfield  {author} {\bibinfo {author} {\bibfnamefont {G.}~\bibnamefont
  {Kresse}}\ and\ \bibinfo {author} {\bibfnamefont {D.}~\bibnamefont
  {Joubert}},\ }\bibfield  {title} {\bibinfo {title} {{From ultrasoft
  pseudopotentials to the projector augmented-wave method}},\ }\href {\doibase
  10.1103/PhysRevB.59.1758} {\bibfield  {journal} {\bibinfo  {journal} {Phys.
  Rev. B}\ }\textbf {\bibinfo {volume} {59}},\ \bibinfo {pages} {1758--1775}
  (\bibinfo {year} {1999})}\BibitemShut {NoStop}%
\bibitem [{\citenamefont {Perdew}\ \emph {et~al.}(1996)\citenamefont {Perdew},
  \citenamefont {Burke},\ and\ \citenamefont
  {Ernzerhof}}]{perdew1996generalized}%
  \BibitemOpen
  \bibfield  {author} {\bibinfo {author} {\bibfnamefont {John~P.}\ \bibnamefont
  {Perdew}}, \bibinfo {author} {\bibfnamefont {Kieron}\ \bibnamefont {Burke}},
  \ and\ \bibinfo {author} {\bibfnamefont {Matthias}\ \bibnamefont
  {Ernzerhof}},\ }\bibfield  {title} {\bibinfo {title} {{Generalized Gradient
  Approximation Made Simple}},\ }\href {\doibase 10.1103/PhysRevLett.77.3865}
  {\bibfield  {journal} {\bibinfo  {journal} {Phys. Rev. Lett.}\ }\textbf
  {\bibinfo {volume} {77}},\ \bibinfo {pages} {3865--3868} (\bibinfo {year}
  {1996})}\BibitemShut {NoStop}%
\bibitem [{\citenamefont {Kresse}\ and\ \citenamefont
  {Furthm{\"u}ller}(1996)}]{kresse1996efficiency}%
  \BibitemOpen
  \bibfield  {author} {\bibinfo {author} {\bibfnamefont {Georg}\ \bibnamefont
  {Kresse}}\ and\ \bibinfo {author} {\bibfnamefont {J{\"u}rgen}\ \bibnamefont
  {Furthm{\"u}ller}},\ }\bibfield  {title} {\bibinfo {title} {{Efficiency of
  ab-initio total energy calculations for metals and semiconductors using a
  plane-wave basis set}},\ }\href
  {https://www.sciencedirect.com/science/article/abs/pii/0927025696000080}
  {\bibfield  {journal} {\bibinfo  {journal} {Comput. Mater. Sci.}\ }\textbf
  {\bibinfo {volume} {6}},\ \bibinfo {pages} {15--50} (\bibinfo {year}
  {1996})}\BibitemShut {NoStop}%
\bibitem [{\citenamefont {Kresse}\ and\ \citenamefont
  {Furthm\"uller}(1996)}]{kresse1996efficient}%
  \BibitemOpen
  \bibfield  {author} {\bibinfo {author} {\bibfnamefont {G.}~\bibnamefont
  {Kresse}}\ and\ \bibinfo {author} {\bibfnamefont {J.}~\bibnamefont
  {Furthm\"uller}},\ }\bibfield  {title} {\bibinfo {title} {{Efficient
  iterative schemes for ab initio total-energy calculations using a plane-wave
  basis set}},\ }\href {\doibase 10.1103/PhysRevB.54.11169} {\bibfield
  {journal} {\bibinfo  {journal} {Phys. Rev. B}\ }\textbf {\bibinfo {volume}
  {54}},\ \bibinfo {pages} {11169--11186} (\bibinfo {year} {1996})}\BibitemShut
  {NoStop}%
\bibitem [{\citenamefont {Monkhorst}\ and\ \citenamefont
  {Pack}(1976)}]{monkhorst1976special}%
  \BibitemOpen
  \bibfield  {author} {\bibinfo {author} {\bibfnamefont {Hendrik~J.}\
  \bibnamefont {Monkhorst}}\ and\ \bibinfo {author} {\bibfnamefont {James~D.}\
  \bibnamefont {Pack}},\ }\bibfield  {title} {\bibinfo {title} {{Special points
  for Brillouin-zone integrations}},\ }\href {\doibase
  10.1103/PhysRevB.13.5188} {\bibfield  {journal} {\bibinfo  {journal} {Phys.
  Rev. B}\ }\textbf {\bibinfo {volume} {13}},\ \bibinfo {pages} {5188--5192}
  (\bibinfo {year} {1976})}\BibitemShut {NoStop}%
\bibitem [{\citenamefont {Coulais}\ \emph {et~al.}(2018)\citenamefont
  {Coulais}, \citenamefont {Sabbadini}, \citenamefont {Vink},\ and\
  \citenamefont {van Hecke}}]{coulais2018multi}%
  \BibitemOpen
  \bibfield  {author} {\bibinfo {author} {\bibfnamefont {Corentin}\
  \bibnamefont {Coulais}}, \bibinfo {author} {\bibfnamefont {Alberico}\
  \bibnamefont {Sabbadini}}, \bibinfo {author} {\bibfnamefont {Fr{\'e}}\
  \bibnamefont {Vink}}, \ and\ \bibinfo {author} {\bibfnamefont {Martin}\
  \bibnamefont {van Hecke}},\ }\bibfield  {title} {\bibinfo {title}
  {{Multi-step self-guided pathways for shape-changing metamaterials}},\ }\href
  {https://doi.org/10.1038/s41586-018-0541-0} {\bibfield  {journal} {\bibinfo
  {journal} {Nature}\ }\textbf {\bibinfo {volume} {561}},\ \bibinfo {pages}
  {512--515} (\bibinfo {year} {2018})}\BibitemShut {NoStop}%
\bibitem [{\citenamefont {Crossno}\ \emph {et~al.}(2016)\citenamefont
  {Crossno}, \citenamefont {Shi}, \citenamefont {Wang}, \citenamefont {Liu},
  \citenamefont {Harzheim}, \citenamefont {Lucas}, \citenamefont {Sachdev},
  \citenamefont {Kim}, \citenamefont {Taniguchi}, \citenamefont {Watanabe},
  \citenamefont {A.~Ohki},\ and\ \citenamefont
  {Chungfong}}]{crossno2016observation}%
  \BibitemOpen
  \bibfield  {author} {\bibinfo {author} {\bibfnamefont {Jesse}\ \bibnamefont
  {Crossno}}, \bibinfo {author} {\bibfnamefont {Jing~K}\ \bibnamefont {Shi}},
  \bibinfo {author} {\bibfnamefont {Ke}~\bibnamefont {Wang}}, \bibinfo {author}
  {\bibfnamefont {Xiaomeng}\ \bibnamefont {Liu}}, \bibinfo {author}
  {\bibfnamefont {Achim}\ \bibnamefont {Harzheim}}, \bibinfo {author}
  {\bibfnamefont {Andrew}\ \bibnamefont {Lucas}}, \bibinfo {author}
  {\bibfnamefont {Subir}\ \bibnamefont {Sachdev}}, \bibinfo {author}
  {\bibfnamefont {Philip}\ \bibnamefont {Kim}}, \bibinfo {author}
  {\bibfnamefont {Takashi}\ \bibnamefont {Taniguchi}}, \bibinfo {author}
  {\bibfnamefont {Kenji}\ \bibnamefont {Watanabe}}, \bibinfo {author}
  {\bibfnamefont {Thomas}\ \bibnamefont {A.~Ohki}}, \ and\ \bibinfo {author}
  {\bibfnamefont {Kin}\ \bibnamefont {Chungfong}},\ }\bibfield  {title}
  {\bibinfo {title} {{Observation of the Dirac fluid and the breakdown of the
  Wiedemann-Franz law in graphene}},\ }\href
  {https://www.science.org/doi/full/10.1126/science.aad0343} {\bibfield
  {journal} {\bibinfo  {journal} {Science}\ }\textbf {\bibinfo {volume}
  {351}},\ \bibinfo {pages} {1058--1061} (\bibinfo {year} {2016})}\BibitemShut
  {NoStop}%
\bibitem [{\citenamefont {Bardeen}\ and\ \citenamefont
  {Shockley}(1950)}]{bardeen1950deformation}%
  \BibitemOpen
  \bibfield  {author} {\bibinfo {author} {\bibfnamefont {J}~\bibnamefont
  {Bardeen}}\ and\ \bibinfo {author} {\bibfnamefont {W}~\bibnamefont
  {Shockley}},\ }\bibfield  {title} {\bibinfo {title} {{Deformation potentials
  and mobilities in non-polar crystals}},\ }\href
  {https://journals.aps.org/pr/abstract/10.1103/PhysRev.80.72} {\bibfield
  {journal} {\bibinfo  {journal} {Phys. Rev. B}\ }\textbf {\bibinfo {volume}
  {80}},\ \bibinfo {pages} {72} (\bibinfo {year} {1950})}\BibitemShut {NoStop}%
\bibitem [{\citenamefont {Xi}\ \emph {et~al.}(2012)\citenamefont {Xi},
  \citenamefont {Long}, \citenamefont {Tang}, \citenamefont {Wang},\ and\
  \citenamefont {Shuai}}]{xi2012first}%
  \BibitemOpen
  \bibfield  {author} {\bibinfo {author} {\bibfnamefont {Jinyang}\ \bibnamefont
  {Xi}}, \bibinfo {author} {\bibfnamefont {Mengqiu}\ \bibnamefont {Long}},
  \bibinfo {author} {\bibfnamefont {Ling}\ \bibnamefont {Tang}}, \bibinfo
  {author} {\bibfnamefont {Dong}\ \bibnamefont {Wang}}, \ and\ \bibinfo
  {author} {\bibfnamefont {Zhigang}\ \bibnamefont {Shuai}},\ }\bibfield
  {title} {\bibinfo {title} {{First-principles prediction of charge mobility in
  carbon and organic nanomaterials}},\ }\href
  {https://pubs.rsc.org/en/content/articlelanding/2012/nr/c2nr30585b/unauth}
  {\bibfield  {journal} {\bibinfo  {journal} {Nanoscale}\ }\textbf {\bibinfo
  {volume} {4}},\ \bibinfo {pages} {4348--4369} (\bibinfo {year}
  {2012})}\BibitemShut {NoStop}%
\bibitem [{\citenamefont {Cai}\ \emph {et~al.}(2014)\citenamefont {Cai},
  \citenamefont {Zhang},\ and\ \citenamefont {Zhang}}]{cai2014polarity}%
  \BibitemOpen
  \bibfield  {author} {\bibinfo {author} {\bibfnamefont {Yongqing}\
  \bibnamefont {Cai}}, \bibinfo {author} {\bibfnamefont {Gang}\ \bibnamefont
  {Zhang}}, \ and\ \bibinfo {author} {\bibfnamefont {Yong-Wei}\ \bibnamefont
  {Zhang}},\ }\bibfield  {title} {\bibinfo {title} {{Polarity-reversed robust
  carrier mobility in monolayer MoS2 nanoribbons}},\ }\href
  {https://pubs.acs.org/doi/abs/10.1021/ja4109787} {\bibfield  {journal}
  {\bibinfo  {journal} {J. Am. Chem. Soc.}\ }\textbf {\bibinfo {volume}
  {136}},\ \bibinfo {pages} {6269--6275} (\bibinfo {year} {2014})}\BibitemShut
  {NoStop}%
\bibitem [{\citenamefont {Yang}\ \emph {et~al.}(2016)\citenamefont {Yang},
  \citenamefont {Zhang}, \citenamefont {Yin}, \citenamefont {Gong},
  \citenamefont {Yakobson},\ and\ \citenamefont {Wei}}]{yang2016two}%
  \BibitemOpen
  \bibfield  {author} {\bibinfo {author} {\bibfnamefont {Ji-Hui}\ \bibnamefont
  {Yang}}, \bibinfo {author} {\bibfnamefont {Yueyu}\ \bibnamefont {Zhang}},
  \bibinfo {author} {\bibfnamefont {Wan-Jian}\ \bibnamefont {Yin}}, \bibinfo
  {author} {\bibfnamefont {XG}~\bibnamefont {Gong}}, \bibinfo {author}
  {\bibfnamefont {Boris~I}\ \bibnamefont {Yakobson}}, \ and\ \bibinfo {author}
  {\bibfnamefont {Su-Huai}\ \bibnamefont {Wei}},\ }\bibfield  {title} {\bibinfo
  {title} {{Two-dimensional SiS layers with promising electronic and
  optoelectronic properties: theoretical prediction}},\ }\href
  {https://pubs.acs.org/doi/abs/10.1021/acs.nanolett.5b04341} {\bibfield
  {journal} {\bibinfo  {journal} {Nano Lett.}\ }\textbf {\bibinfo {volume}
  {16}},\ \bibinfo {pages} {1110--1117} (\bibinfo {year} {2016})}\BibitemShut
  {NoStop}%
\bibitem [{\citenamefont {Zhu}\ \emph {et~al.}(2017)\citenamefont {Zhu},
  \citenamefont {Cai}, \citenamefont {Yi}, \citenamefont {Chen}, \citenamefont
  {Dai}, \citenamefont {Niu}, \citenamefont {Guo}, \citenamefont {Xie},
  \citenamefont {Liu}, \citenamefont {Cho}, \citenamefont {Jia},\ and\
  \citenamefont {Zhang}}]{zhu2017multivalency}%
  \BibitemOpen
  \bibfield  {author} {\bibinfo {author} {\bibfnamefont {Zhili}\ \bibnamefont
  {Zhu}}, \bibinfo {author} {\bibfnamefont {Xiaolin}\ \bibnamefont {Cai}},
  \bibinfo {author} {\bibfnamefont {Seho}\ \bibnamefont {Yi}}, \bibinfo
  {author} {\bibfnamefont {Jinglei}\ \bibnamefont {Chen}}, \bibinfo {author}
  {\bibfnamefont {Yawei}\ \bibnamefont {Dai}}, \bibinfo {author} {\bibfnamefont
  {Chunyao}\ \bibnamefont {Niu}}, \bibinfo {author} {\bibfnamefont {Zhengxiao}\
  \bibnamefont {Guo}}, \bibinfo {author} {\bibfnamefont {Maohai}\ \bibnamefont
  {Xie}}, \bibinfo {author} {\bibfnamefont {Feng}\ \bibnamefont {Liu}},
  \bibinfo {author} {\bibfnamefont {Jun-Hyung}\ \bibnamefont {Cho}}, \bibinfo
  {author} {\bibfnamefont {Yu}~\bibnamefont {Jia}}, \ and\ \bibinfo {author}
  {\bibfnamefont {Zhenyu}\ \bibnamefont {Zhang}},\ }\bibfield  {title}
  {\bibinfo {title} {{Multivalency-driven formation of Te-based monolayer
  materials: a combined first-principles and experimental study}},\ }\href
  {https://journals.aps.org/prl/abstract/10.1103/PhysRevLett.119.106101}
  {\bibfield  {journal} {\bibinfo  {journal} {Phys. Rev. Lett.}\ }\textbf
  {\bibinfo {volume} {119}},\ \bibinfo {pages} {106101} (\bibinfo {year}
  {2017})}\BibitemShut {NoStop}%
\bibitem [{\citenamefont {Plimpton}(1995)}]{plimpton1995fast}%
  \BibitemOpen
  \bibfield  {author} {\bibinfo {author} {\bibfnamefont {Steve}\ \bibnamefont
  {Plimpton}},\ }\bibfield  {title} {\bibinfo {title} {{Fast parallel
  algorithms for short-range molecular dynamics}},\ }\href
  {https://www.sciencedirect.com/science/article/pii/S002199918571039X}
  {\bibfield  {journal} {\bibinfo  {journal} {J. Comput. Phys.}\ }\textbf
  {\bibinfo {volume} {117}},\ \bibinfo {pages} {1--19} (\bibinfo {year}
  {1995})}\BibitemShut {NoStop}%
\bibitem [{\citenamefont {Kinaci}\ \emph {et~al.}(2012)\citenamefont {Kinaci},
  \citenamefont {Haskins}, \citenamefont {Sevik},\ and\ \citenamefont {\ifmmode
  \mbox{\c{C}}\else \c{C}\fi{}a\ifmmode~\breve{g}\else
  \u{g}\fi{}in}}]{kinaci2012thermal}%
  \BibitemOpen
  \bibfield  {author} {\bibinfo {author} {\bibfnamefont {Alper}\ \bibnamefont
  {Kinaci}}, \bibinfo {author} {\bibfnamefont {Justin~B.}\ \bibnamefont
  {Haskins}}, \bibinfo {author} {\bibfnamefont {Cem}\ \bibnamefont {Sevik}}, \
  and\ \bibinfo {author} {\bibfnamefont {Tahir}\ \bibnamefont {\ifmmode
  \mbox{\c{C}}\else \c{C}\fi{}a\ifmmode~\breve{g}\else \u{g}\fi{}in}},\
  }\bibfield  {title} {\bibinfo {title} {{Thermal conductivity of BN-C
  nanostructures}},\ }\href {\doibase 10.1103/PhysRevB.86.115410} {\bibfield
  {journal} {\bibinfo  {journal} {Phys. Rev. B}\ }\textbf {\bibinfo {volume}
  {86}},\ \bibinfo {pages} {115410} (\bibinfo {year} {2012})}\BibitemShut
  {NoStop}%
\bibitem [{\citenamefont {Xu}\ \emph {et~al.}(2014)\citenamefont {Xu},
  \citenamefont {Pereira}, \citenamefont {Wang}, \citenamefont {Wu},
  \citenamefont {Zhang}, \citenamefont {Zhao}, \citenamefont {Bae},
  \citenamefont {Bui}, \citenamefont {Xie}, \citenamefont {Thong},
  \citenamefont {Hong}, \citenamefont {Loh}, \citenamefont {Donadio},
  \citenamefont {Li},\ and\ \citenamefont {Özyilmaz}}]{xu2014length}%
  \BibitemOpen
  \bibfield  {author} {\bibinfo {author} {\bibfnamefont {Xiangfan}\
  \bibnamefont {Xu}}, \bibinfo {author} {\bibfnamefont {Luiz~FC}\ \bibnamefont
  {Pereira}}, \bibinfo {author} {\bibfnamefont {Yu}~\bibnamefont {Wang}},
  \bibinfo {author} {\bibfnamefont {Jing}\ \bibnamefont {Wu}}, \bibinfo
  {author} {\bibfnamefont {Kaiwen}\ \bibnamefont {Zhang}}, \bibinfo {author}
  {\bibfnamefont {Xiangming}\ \bibnamefont {Zhao}}, \bibinfo {author}
  {\bibfnamefont {Sukang}\ \bibnamefont {Bae}}, \bibinfo {author}
  {\bibfnamefont {Cong~Tinh}\ \bibnamefont {Bui}}, \bibinfo {author}
  {\bibfnamefont {Rongguo}\ \bibnamefont {Xie}}, \bibinfo {author}
  {\bibfnamefont {John~TL}\ \bibnamefont {Thong}}, \bibinfo {author}
  {\bibfnamefont {Byung~Hee}\ \bibnamefont {Hong}}, \bibinfo {author}
  {\bibfnamefont {Kian~Ping}\ \bibnamefont {Loh}}, \bibinfo {author}
  {\bibfnamefont {Davide}\ \bibnamefont {Donadio}}, \bibinfo {author}
  {\bibfnamefont {Baowen}\ \bibnamefont {Li}}, \ and\ \bibinfo {author}
  {\bibfnamefont {Barbaros}\ \bibnamefont {Özyilmaz}},\ }\bibfield  {title}
  {\bibinfo {title} {{Length-dependent thermal conductivity in suspended
  single-layer graphene}},\ }\href {https://doi.org/10.1038/ncomms4689}
  {\bibfield  {journal} {\bibinfo  {journal} {Nat. Commun.}\ }\textbf {\bibinfo
  {volume} {5}},\ \bibinfo {pages} {3689} (\bibinfo {year} {2014})}\BibitemShut
  {NoStop}%
\bibitem [{\citenamefont {Wang}\ \emph {et~al.}(2017)\citenamefont {Wang},
  \citenamefont {Hu}, \citenamefont {Takahashi}, \citenamefont {Zhang},
  \citenamefont {Takamatsu},\ and\ \citenamefont
  {Chen}}]{wang2017experimental}%
  \BibitemOpen
  \bibfield  {author} {\bibinfo {author} {\bibfnamefont {Haidong}\ \bibnamefont
  {Wang}}, \bibinfo {author} {\bibfnamefont {Shiqian}\ \bibnamefont {Hu}},
  \bibinfo {author} {\bibfnamefont {Koji}\ \bibnamefont {Takahashi}}, \bibinfo
  {author} {\bibfnamefont {Xing}\ \bibnamefont {Zhang}}, \bibinfo {author}
  {\bibfnamefont {Hiroshi}\ \bibnamefont {Takamatsu}}, \ and\ \bibinfo {author}
  {\bibfnamefont {Jie}\ \bibnamefont {Chen}},\ }\bibfield  {title} {\bibinfo
  {title} {{Experimental study of thermal rectification in suspended monolayer
  graphene}},\ }\href {https://doi.org/10.1038/ncomms15843} {\bibfield
  {journal} {\bibinfo  {journal} {Nat. Commun.}\ }\textbf {\bibinfo {volume}
  {8}},\ \bibinfo {pages} {15843} (\bibinfo {year} {2017})}\BibitemShut
  {NoStop}%
\bibitem [{\citenamefont {Lepri}\ \emph {et~al.}(1997)\citenamefont {Lepri},
  \citenamefont {Livi},\ and\ \citenamefont {Politi}}]{lepri1997heat}%
  \BibitemOpen
  \bibfield  {author} {\bibinfo {author} {\bibfnamefont {Stefano}\ \bibnamefont
  {Lepri}}, \bibinfo {author} {\bibfnamefont {Roberto}\ \bibnamefont {Livi}}, \
  and\ \bibinfo {author} {\bibfnamefont {Antonio}\ \bibnamefont {Politi}},\
  }\bibfield  {title} {\bibinfo {title} {{Heat Conduction in Chains of
  Nonlinear Oscillators}},\ }\href {\doibase 10.1103/PhysRevLett.78.1896}
  {\bibfield  {journal} {\bibinfo  {journal} {Phys. Rev. Lett.}\ }\textbf
  {\bibinfo {volume} {78}},\ \bibinfo {pages} {1896--1899} (\bibinfo {year}
  {1997})}\BibitemShut {NoStop}%
\bibitem [{\citenamefont {Gao}\ \emph {et~al.}(2016)\citenamefont {Gao},
  \citenamefont {Li},\ and\ \citenamefont {Li}}]{gao2016heat}%
  \BibitemOpen
  \bibfield  {author} {\bibinfo {author} {\bibfnamefont {Zhibin}\ \bibnamefont
  {Gao}}, \bibinfo {author} {\bibfnamefont {Nianbei}\ \bibnamefont {Li}}, \
  and\ \bibinfo {author} {\bibfnamefont {Baowen}\ \bibnamefont {Li}},\
  }\bibfield  {title} {\bibinfo {title} {{Heat conduction and energy diffusion
  in momentum-conserving one-dimensional full-lattice ding-a-ling model}},\
  }\href {\doibase 10.1103/PhysRevE.93.022102} {\bibfield  {journal} {\bibinfo
  {journal} {Phys. Rev. E}\ }\textbf {\bibinfo {volume} {93}},\ \bibinfo
  {pages} {022102} (\bibinfo {year} {2016})}\BibitemShut {NoStop}%
\bibitem [{\citenamefont {Berglund}\ and\ \citenamefont
  {Guggenheim}(1969)}]{berglund1969electronic}%
  \BibitemOpen
  \bibfield  {author} {\bibinfo {author} {\bibfnamefont {C.~N.}\ \bibnamefont
  {Berglund}}\ and\ \bibinfo {author} {\bibfnamefont {H.~J.}\ \bibnamefont
  {Guggenheim}},\ }\bibfield  {title} {\bibinfo {title} {{Electronic Properties
  of V${\mathrm{O}}_{2}$ near the Semiconductor-Metal Transition}},\ }\href
  {\doibase 10.1103/PhysRev.185.1022} {\bibfield  {journal} {\bibinfo
  {journal} {Phys. Rev.}\ }\textbf {\bibinfo {volume} {185}},\ \bibinfo {pages}
  {1022--1033} (\bibinfo {year} {1969})}\BibitemShut {NoStop}%
\bibitem [{\citenamefont {Naumis}\ \emph {et~al.}(2017)\citenamefont {Naumis},
  \citenamefont {Barraza-Lopez}, \citenamefont {Oliva-Leyva},\ and\
  \citenamefont {Terrones}}]{naumis2017electronic}%
  \BibitemOpen
  \bibfield  {author} {\bibinfo {author} {\bibfnamefont {Gerardo~G}\
  \bibnamefont {Naumis}}, \bibinfo {author} {\bibfnamefont {Salvador}\
  \bibnamefont {Barraza-Lopez}}, \bibinfo {author} {\bibfnamefont {Maurice}\
  \bibnamefont {Oliva-Leyva}}, \ and\ \bibinfo {author} {\bibfnamefont
  {Humberto}\ \bibnamefont {Terrones}},\ }\bibfield  {title} {\bibinfo {title}
  {{Electronic and optical properties of strained graphene and other strained
  2D materials: a review}},\ }\href
  {https://iopscience.iop.org/article/10.1088/1361-6633/aa74ef/meta} {\bibfield
   {journal} {\bibinfo  {journal} {Rep. Prog. Phys.}\ }\textbf {\bibinfo
  {volume} {80}},\ \bibinfo {pages} {096501} (\bibinfo {year}
  {2017})}\BibitemShut {NoStop}%
\bibitem [{\citenamefont {Ghorbani-Asl}\ \emph {et~al.}(2013)\citenamefont
  {Ghorbani-Asl}, \citenamefont {Borini}, \citenamefont {Kuc},\ and\
  \citenamefont {Heine}}]{ghorbani2013strain}%
  \BibitemOpen
  \bibfield  {author} {\bibinfo {author} {\bibfnamefont {M.}~\bibnamefont
  {Ghorbani-Asl}}, \bibinfo {author} {\bibfnamefont {S.}~\bibnamefont
  {Borini}}, \bibinfo {author} {\bibfnamefont {A.}~\bibnamefont {Kuc}}, \ and\
  \bibinfo {author} {\bibfnamefont {T.}~\bibnamefont {Heine}},\ }\bibfield
  {title} {\bibinfo {title} {{Strain-dependent modulation of conductivity in
  single-layer transition-metal dichalcogenides}},\ }\href {\doibase
  10.1103/PhysRevB.87.235434} {\bibfield  {journal} {\bibinfo  {journal} {Phys.
  Rev. B}\ }\textbf {\bibinfo {volume} {87}},\ \bibinfo {pages} {235434}
  (\bibinfo {year} {2013})}\BibitemShut {NoStop}%
\bibitem [{\citenamefont {Chang}\ \emph {et~al.}(2013)\citenamefont {Chang},
  \citenamefont {Fan}, \citenamefont {Lin},\ and\ \citenamefont
  {Kuo}}]{chang2013orbital}%
  \BibitemOpen
  \bibfield  {author} {\bibinfo {author} {\bibfnamefont {Chung-Huai}\
  \bibnamefont {Chang}}, \bibinfo {author} {\bibfnamefont {Xiaofeng}\
  \bibnamefont {Fan}}, \bibinfo {author} {\bibfnamefont {Shi-Hsin}\
  \bibnamefont {Lin}}, \ and\ \bibinfo {author} {\bibfnamefont {Jer-Lai}\
  \bibnamefont {Kuo}},\ }\bibfield  {title} {\bibinfo {title} {{Orbital
  analysis of electronic structure and phonon dispersion in MoS${}_{2}$,
  MoSe${}_{2}$, WS${}_{2}$, and WSe${}_{2}$ monolayers under strain}},\ }\href
  {\doibase 10.1103/PhysRevB.88.195420} {\bibfield  {journal} {\bibinfo
  {journal} {Phys. Rev. B}\ }\textbf {\bibinfo {volume} {88}},\ \bibinfo
  {pages} {195420} (\bibinfo {year} {2013})}\BibitemShut {NoStop}%
\bibitem [{\citenamefont {Scalise}\ \emph {et~al.}(2012)\citenamefont
  {Scalise}, \citenamefont {Houssa}, \citenamefont {Pourtois}, \citenamefont
  {Afanas’ev},\ and\ \citenamefont {Stesmans}}]{scalise2012strain}%
  \BibitemOpen
  \bibfield  {author} {\bibinfo {author} {\bibfnamefont {Emilio}\ \bibnamefont
  {Scalise}}, \bibinfo {author} {\bibfnamefont {Michel}\ \bibnamefont
  {Houssa}}, \bibinfo {author} {\bibfnamefont {Geoffrey}\ \bibnamefont
  {Pourtois}}, \bibinfo {author} {\bibfnamefont {Valery}\ \bibnamefont
  {Afanas’ev}}, \ and\ \bibinfo {author} {\bibfnamefont {Andr{\'e}}\
  \bibnamefont {Stesmans}},\ }\bibfield  {title} {\bibinfo {title}
  {{Strain-induced semiconductor to metal transition in the two-dimensional
  honeycomb structure of MoS${}_{2}$}},\ }\href
  {https://link.springer.com/article/10.1007/s12274-011-0183-0} {\bibfield
  {journal} {\bibinfo  {journal} {Nano Res.}\ }\textbf {\bibinfo {volume}
  {5}},\ \bibinfo {pages} {43--48} (\bibinfo {year} {2012})}\BibitemShut
  {NoStop}%
\bibitem [{\citenamefont {Brandbyge}\ \emph {et~al.}(2002)\citenamefont
  {Brandbyge}, \citenamefont {Mozos}, \citenamefont {Ordej\'on}, \citenamefont
  {Taylor},\ and\ \citenamefont {Stokbro}}]{brandbyge2002density}%
  \BibitemOpen
  \bibfield  {author} {\bibinfo {author} {\bibfnamefont {Mads}\ \bibnamefont
  {Brandbyge}}, \bibinfo {author} {\bibfnamefont {Jos\'e-Luis}\ \bibnamefont
  {Mozos}}, \bibinfo {author} {\bibfnamefont {Pablo}\ \bibnamefont
  {Ordej\'on}}, \bibinfo {author} {\bibfnamefont {Jeremy}\ \bibnamefont
  {Taylor}}, \ and\ \bibinfo {author} {\bibfnamefont {Kurt}\ \bibnamefont
  {Stokbro}},\ }\bibfield  {title} {\bibinfo {title} {{Density-functional
  method for nonequilibrium electron transport}},\ }\href {\doibase
  10.1103/PhysRevB.65.165401} {\bibfield  {journal} {\bibinfo  {journal} {Phys.
  Rev. B}\ }\textbf {\bibinfo {volume} {65}},\ \bibinfo {pages} {165401}
  (\bibinfo {year} {2002})}\BibitemShut {NoStop}%
\bibitem [{\citenamefont {Smidstrup}\ \emph {et~al.}(2019)\citenamefont
  {Smidstrup}, \citenamefont {Markussen}, \citenamefont {Vancraeyveld},
  \citenamefont {Wellendorff}, \citenamefont {Schneider}, \citenamefont
  {Gunst}, \citenamefont {Verstichel}, \citenamefont {Stradi}, \citenamefont
  {Khomyakov}, \citenamefont {Vej-Hansen}, \citenamefont {Lee}, \citenamefont
  {Chill}, \citenamefont {Rasmussen}, \citenamefont {Penazzi}, \citenamefont
  {Corsetti}, \citenamefont {Ojanperä}, \citenamefont {Jensen}, \citenamefont
  {Palsgaard}, \citenamefont {Martinez}, \citenamefont {Blom}, \citenamefont
  {Brandbyge},\ and\ \citenamefont {Stokbro}}]{smidstrup2019quantumatk}%
  \BibitemOpen
  \bibfield  {author} {\bibinfo {author} {\bibfnamefont {S{\o}ren}\
  \bibnamefont {Smidstrup}}, \bibinfo {author} {\bibfnamefont {Troels}\
  \bibnamefont {Markussen}}, \bibinfo {author} {\bibfnamefont {Pieter}\
  \bibnamefont {Vancraeyveld}}, \bibinfo {author} {\bibfnamefont {Jess}\
  \bibnamefont {Wellendorff}}, \bibinfo {author} {\bibfnamefont {Julian}\
  \bibnamefont {Schneider}}, \bibinfo {author} {\bibfnamefont {Tue}\
  \bibnamefont {Gunst}}, \bibinfo {author} {\bibfnamefont {Brecht}\
  \bibnamefont {Verstichel}}, \bibinfo {author} {\bibfnamefont {Daniele}\
  \bibnamefont {Stradi}}, \bibinfo {author} {\bibfnamefont {Petr~A}\
  \bibnamefont {Khomyakov}}, \bibinfo {author} {\bibfnamefont {Ulrik~G}\
  \bibnamefont {Vej-Hansen}}, \bibinfo {author} {\bibfnamefont {Maeng-Eun}\
  \bibnamefont {Lee}}, \bibinfo {author} {\bibfnamefont {Samuel~T}\
  \bibnamefont {Chill}}, \bibinfo {author} {\bibfnamefont {Filip}\ \bibnamefont
  {Rasmussen}}, \bibinfo {author} {\bibfnamefont {Gabriele}\ \bibnamefont
  {Penazzi}}, \bibinfo {author} {\bibfnamefont {Fabiano}\ \bibnamefont
  {Corsetti}}, \bibinfo {author} {\bibfnamefont {Ari}\ \bibnamefont
  {Ojanperä}}, \bibinfo {author} {\bibfnamefont {Kristian}\ \bibnamefont
  {Jensen}}, \bibinfo {author} {\bibfnamefont {Mattias L~N}\ \bibnamefont
  {Palsgaard}}, \bibinfo {author} {\bibfnamefont {Umberto}\ \bibnamefont
  {Martinez}}, \bibinfo {author} {\bibfnamefont {Anders}\ \bibnamefont {Blom}},
  \bibinfo {author} {\bibfnamefont {Mads}\ \bibnamefont {Brandbyge}}, \ and\
  \bibinfo {author} {\bibfnamefont {Kurt}\ \bibnamefont {Stokbro}},\ }\bibfield
   {title} {\bibinfo {title} {{QuantumATK: an integrated platform of electronic
  and atomic-scale modelling tools}},\ }\href
  {https://iopscience.iop.org/article/10.1088/1361-648X/ab4007/meta} {\bibfield
   {journal} {\bibinfo  {journal} {J. Phys.: Condens. Matter.}\ }\textbf
  {\bibinfo {volume} {32}},\ \bibinfo {pages} {015901} (\bibinfo {year}
  {2019})}\BibitemShut {NoStop}%
\bibitem [{\citenamefont {Datta}(1996)}]{datta1996electronic}%
  \BibitemOpen
  \bibfield  {author} {\bibinfo {author} {\bibfnamefont {Supriyo}\ \bibnamefont
  {Datta}},\ }\bibfield  {title} {\bibinfo {title} {{Electronic transport in
  mesoscopic systems}},\ }\href {https://doi.org/10.1063/1.2807624} {\bibfield
  {journal} {\bibinfo  {journal} {Phys. Today}\ }\textbf {\bibinfo {volume}
  {49}},\ \bibinfo {pages} {70} (\bibinfo {year} {1996})}\BibitemShut {NoStop}%
\bibitem [{\citenamefont {Liu}\ \emph {et~al.}(2019{\natexlab{b}})\citenamefont
  {Liu}, \citenamefont {Gao},\ and\ \citenamefont {Ren}}]{liu2019anisotropic}%
  \BibitemOpen
  \bibfield  {author} {\bibinfo {author} {\bibfnamefont {Gang}\ \bibnamefont
  {Liu}}, \bibinfo {author} {\bibfnamefont {Zhibin}\ \bibnamefont {Gao}}, \
  and\ \bibinfo {author} {\bibfnamefont {Jie}\ \bibnamefont {Ren}},\ }\bibfield
   {title} {\bibinfo {title} {{Anisotropic thermal expansion and thermodynamic
  properties of monolayer $\ensuremath{\beta}$-Te}},\ }\href {\doibase
  10.1103/PhysRevB.99.195436} {\bibfield  {journal} {\bibinfo  {journal} {Phys.
  Rev. B}\ }\textbf {\bibinfo {volume} {99}},\ \bibinfo {pages} {195436}
  (\bibinfo {year} {2019}{\natexlab{b}})}\BibitemShut {NoStop}%
\bibitem [{\citenamefont {Yoon}\ \emph {et~al.}(2011)\citenamefont {Yoon},
  \citenamefont {Son},\ and\ \citenamefont {Cheong}}]{yoon2011negative}%
  \BibitemOpen
  \bibfield  {author} {\bibinfo {author} {\bibfnamefont {Duhee}\ \bibnamefont
  {Yoon}}, \bibinfo {author} {\bibfnamefont {Young-Woo}\ \bibnamefont {Son}}, \
  and\ \bibinfo {author} {\bibfnamefont {Hyeonsik}\ \bibnamefont {Cheong}},\
  }\bibfield  {title} {\bibinfo {title} {{Negative thermal expansion
  coefficient of graphene measured by Raman spectroscopy}},\ }\href
  {https://doi.org/10.1021/nl201488g} {\bibfield  {journal} {\bibinfo
  {journal} {Nano Lett.}\ }\textbf {\bibinfo {volume} {11}},\ \bibinfo {pages}
  {3227--3231} (\bibinfo {year} {2011})}\BibitemShut {NoStop}%
\bibitem [{\citenamefont {Liu}\ \emph {et~al.}(2016)\citenamefont {Liu},
  \citenamefont {Every},\ and\ \citenamefont {Tom{\'a}nek}}]{liu2016continuum}%
  \BibitemOpen
  \bibfield  {author} {\bibinfo {author} {\bibfnamefont {Dan}\ \bibnamefont
  {Liu}}, \bibinfo {author} {\bibfnamefont {Arthur~G}\ \bibnamefont {Every}}, \
  and\ \bibinfo {author} {\bibfnamefont {David}\ \bibnamefont {Tom{\'a}nek}},\
  }\bibfield  {title} {\bibinfo {title} {{Continuum approach for
  long-wavelength acoustic phonons in quasi-two-dimensional structures}},\
  }\href {https://journals.aps.org/prb/abstract/10.1103/PhysRevB.94.165432}
  {\bibfield  {journal} {\bibinfo  {journal} {Phys. Rev. B}\ }\textbf {\bibinfo
  {volume} {94}},\ \bibinfo {pages} {165432} (\bibinfo {year}
  {2016})}\BibitemShut {NoStop}%
\bibitem [{\citenamefont {Kambe}\ \emph {et~al.}(2013)\citenamefont {Kambe},
  \citenamefont {Sakamoto}, \citenamefont {Hoshiko}, \citenamefont {Takada},
  \citenamefont {Miyachi}, \citenamefont {Ryu}, \citenamefont {Sasaki},
  \citenamefont {Kim}, \citenamefont {Nakazato}, \citenamefont {Takata},\ and\
  \citenamefont {Nishihara}}]{kambe2013pi}%
  \BibitemOpen
  \bibfield  {author} {\bibinfo {author} {\bibfnamefont {Tetsuya}\ \bibnamefont
  {Kambe}}, \bibinfo {author} {\bibfnamefont {Ryota}\ \bibnamefont {Sakamoto}},
  \bibinfo {author} {\bibfnamefont {Ken}\ \bibnamefont {Hoshiko}}, \bibinfo
  {author} {\bibfnamefont {Kenji}\ \bibnamefont {Takada}}, \bibinfo {author}
  {\bibfnamefont {Mariko}\ \bibnamefont {Miyachi}}, \bibinfo {author}
  {\bibfnamefont {Ji-Heun}\ \bibnamefont {Ryu}}, \bibinfo {author}
  {\bibfnamefont {Sono}\ \bibnamefont {Sasaki}}, \bibinfo {author}
  {\bibfnamefont {Jungeun}\ \bibnamefont {Kim}}, \bibinfo {author}
  {\bibfnamefont {Kazuo}\ \bibnamefont {Nakazato}}, \bibinfo {author}
  {\bibfnamefont {Masaki}\ \bibnamefont {Takata}}, \ and\ \bibinfo {author}
  {\bibfnamefont {Hiroshi}\ \bibnamefont {Nishihara}},\ }\bibfield  {title}
  {\bibinfo {title} {{$\pi$-Conjugated nickel bis (dithiolene) complex
  nanosheet}},\ }\href {https://doi.org/10.1021/ja312380b} {\bibfield
  {journal} {\bibinfo  {journal} {J. Am. Chem. Soc.}\ }\textbf {\bibinfo
  {volume} {135}},\ \bibinfo {pages} {2462--2465} (\bibinfo {year}
  {2013})}\BibitemShut {NoStop}%
\bibitem [{\citenamefont {Sakamoto}\ \emph {et~al.}(2017)\citenamefont
  {Sakamoto}, \citenamefont {Takada}, \citenamefont {Pal}, \citenamefont
  {Maeda}, \citenamefont {Kambe},\ and\ \citenamefont
  {Nishihara}}]{sakamoto2017coordination}%
  \BibitemOpen
  \bibfield  {author} {\bibinfo {author} {\bibfnamefont {Ryota}\ \bibnamefont
  {Sakamoto}}, \bibinfo {author} {\bibfnamefont {Kenji}\ \bibnamefont
  {Takada}}, \bibinfo {author} {\bibfnamefont {Tigmansu}\ \bibnamefont {Pal}},
  \bibinfo {author} {\bibfnamefont {Hiroaki}\ \bibnamefont {Maeda}}, \bibinfo
  {author} {\bibfnamefont {Tetsuya}\ \bibnamefont {Kambe}}, \ and\ \bibinfo
  {author} {\bibfnamefont {Hiroshi}\ \bibnamefont {Nishihara}},\ }\bibfield
  {title} {\bibinfo {title} {{Coordination nanosheets (CONASHs): strategies,
  structures and functions}},\ }\href {https://doi.org/10.1039/C7CC00810D}
  {\bibfield  {journal} {\bibinfo  {journal} {Chem. Commun.}\ }\textbf
  {\bibinfo {volume} {53}},\ \bibinfo {pages} {5781--5801} (\bibinfo {year}
  {2017})}\BibitemShut {NoStop}%
\bibitem [{\citenamefont {Treier}\ \emph {et~al.}(2011)\citenamefont {Treier},
  \citenamefont {Pignedoli}, \citenamefont {Laino}, \citenamefont {Rieger},
  \citenamefont {M{\"u}llen}, \citenamefont {Passerone},\ and\ \citenamefont
  {Fasel}}]{treier2011surface}%
  \BibitemOpen
  \bibfield  {author} {\bibinfo {author} {\bibfnamefont {Matthias}\
  \bibnamefont {Treier}}, \bibinfo {author} {\bibfnamefont {Carlo~Antonio}\
  \bibnamefont {Pignedoli}}, \bibinfo {author} {\bibfnamefont {Teodoro}\
  \bibnamefont {Laino}}, \bibinfo {author} {\bibfnamefont {Ralph}\ \bibnamefont
  {Rieger}}, \bibinfo {author} {\bibfnamefont {Klaus}\ \bibnamefont
  {M{\"u}llen}}, \bibinfo {author} {\bibfnamefont {Daniele}\ \bibnamefont
  {Passerone}}, \ and\ \bibinfo {author} {\bibfnamefont {Roman}\ \bibnamefont
  {Fasel}},\ }\bibfield  {title} {\bibinfo {title} {{Surface-assisted
  cyclodehydrogenation provides a synthetic route towards easily processable
  and chemically tailored nanographenes}},\ }\href
  {https://doi.org/10.1038/nchem.891} {\bibfield  {journal} {\bibinfo
  {journal} {Nat. Chem.}\ }\textbf {\bibinfo {volume} {3}},\ \bibinfo {pages}
  {61--67} (\bibinfo {year} {2011})}\BibitemShut {NoStop}%
\bibitem [{\citenamefont {Moreno}\ \emph {et~al.}(2018)\citenamefont {Moreno},
  \citenamefont {Vilas-Varela}, \citenamefont {Kretz}, \citenamefont
  {Garcia-Lekue}, \citenamefont {Costache}, \citenamefont {Paradinas},
  \citenamefont {Panighel}, \citenamefont {Ceballos}, \citenamefont
  {Valenzuela}, \citenamefont {Peña},\ and\ \citenamefont
  {Mugarza}}]{moreno2018bottom}%
  \BibitemOpen
  \bibfield  {author} {\bibinfo {author} {\bibfnamefont {César}\ \bibnamefont
  {Moreno}}, \bibinfo {author} {\bibfnamefont {Manuel}\ \bibnamefont
  {Vilas-Varela}}, \bibinfo {author} {\bibfnamefont {Bernhard}\ \bibnamefont
  {Kretz}}, \bibinfo {author} {\bibfnamefont {Aran}\ \bibnamefont
  {Garcia-Lekue}}, \bibinfo {author} {\bibfnamefont {Marius~V.}\ \bibnamefont
  {Costache}}, \bibinfo {author} {\bibfnamefont {Markos}\ \bibnamefont
  {Paradinas}}, \bibinfo {author} {\bibfnamefont {Mirko}\ \bibnamefont
  {Panighel}}, \bibinfo {author} {\bibfnamefont {Gustavo}\ \bibnamefont
  {Ceballos}}, \bibinfo {author} {\bibfnamefont {Sergio~O.}\ \bibnamefont
  {Valenzuela}}, \bibinfo {author} {\bibfnamefont {Diego}\ \bibnamefont
  {Peña}}, \ and\ \bibinfo {author} {\bibfnamefont {Aitor}\ \bibnamefont
  {Mugarza}},\ }\bibfield  {title} {\bibinfo {title} {{Bottom-up synthesis of
  multifunctional nanoporous graphene}},\ }\href {\doibase
  10.1126/science.aar2009} {\bibfield  {journal} {\bibinfo  {journal}
  {Science}\ }\textbf {\bibinfo {volume} {360}},\ \bibinfo {pages} {199--203}
  (\bibinfo {year} {2018})}\BibitemShut {NoStop}%
\bibitem [{\citenamefont {Sugimoto}\ \emph {et~al.}(2015)\citenamefont
  {Sugimoto}, \citenamefont {Fujii},\ and\ \citenamefont
  {Imakita}}]{sugimoto2015size}%
  \BibitemOpen
  \bibfield  {author} {\bibinfo {author} {\bibfnamefont {Hiroshi}\ \bibnamefont
  {Sugimoto}}, \bibinfo {author} {\bibfnamefont {Minoru}\ \bibnamefont
  {Fujii}}, \ and\ \bibinfo {author} {\bibfnamefont {Kenji}\ \bibnamefont
  {Imakita}},\ }\bibfield  {title} {\bibinfo {title} {{Size-controlled growth
  of cubic boron phosphide nanocrystals}},\ }\href
  {https://pubs.rsc.org/en/content/articlehtml/2015/ra/c4ra13530j} {\bibfield
  {journal} {\bibinfo  {journal} {RSC Adv.}\ }\textbf {\bibinfo {volume} {5}},\
  \bibinfo {pages} {8427--8431} (\bibinfo {year} {2015})}\BibitemShut {NoStop}%
\bibitem [{\citenamefont {Liu}\ \emph {et~al.}(2018)\citenamefont {Liu},
  \citenamefont {He}, \citenamefont {Xue}, \citenamefont {Li}, \citenamefont
  {Liu},\ and\ \citenamefont {Edgar}}]{liu2018single}%
  \BibitemOpen
  \bibfield  {author} {\bibinfo {author} {\bibfnamefont {Song}\ \bibnamefont
  {Liu}}, \bibinfo {author} {\bibfnamefont {Rui}\ \bibnamefont {He}}, \bibinfo
  {author} {\bibfnamefont {Lianjie}\ \bibnamefont {Xue}}, \bibinfo {author}
  {\bibfnamefont {Jiahan}\ \bibnamefont {Li}}, \bibinfo {author} {\bibfnamefont
  {Bin}\ \bibnamefont {Liu}}, \ and\ \bibinfo {author} {\bibfnamefont
  {James~H}\ \bibnamefont {Edgar}},\ }\bibfield  {title} {\bibinfo {title}
  {{Single crystal growth of millimeter-sized monoisotopic hexagonal boron
  nitride}},\ }\href
  {https://pubs.acs.org/doi/abs/10.1021/acs.chemmater.8b02589} {\bibfield
  {journal} {\bibinfo  {journal} {Chem. Mater.}\ }\textbf {\bibinfo {volume}
  {30}},\ \bibinfo {pages} {6222--6225} (\bibinfo {year} {2018})}\BibitemShut
  {NoStop}%
\bibitem [{\citenamefont {Bieri}\ \emph {et~al.}(2009)\citenamefont {Bieri},
  \citenamefont {Treier}, \citenamefont {Cai}, \citenamefont {Aït-Mansour},
  \citenamefont {Ruffieux}, \citenamefont {Gröning}, \citenamefont {Gröning},
  \citenamefont {Kastler}, \citenamefont {Rieger}, \citenamefont {Feng},
  \citenamefont {Müllen},\ and\ \citenamefont {Fasel}}]{bieri2009porous}%
  \BibitemOpen
  \bibfield  {author} {\bibinfo {author} {\bibfnamefont {Marco}\ \bibnamefont
  {Bieri}}, \bibinfo {author} {\bibfnamefont {Matthias}\ \bibnamefont
  {Treier}}, \bibinfo {author} {\bibfnamefont {Jinming}\ \bibnamefont {Cai}},
  \bibinfo {author} {\bibfnamefont {Kamel}\ \bibnamefont {Aït-Mansour}},
  \bibinfo {author} {\bibfnamefont {Pascal}\ \bibnamefont {Ruffieux}}, \bibinfo
  {author} {\bibfnamefont {Oliver}\ \bibnamefont {Gröning}}, \bibinfo {author}
  {\bibfnamefont {Pierangelo}\ \bibnamefont {Gröning}}, \bibinfo {author}
  {\bibfnamefont {Marcel}\ \bibnamefont {Kastler}}, \bibinfo {author}
  {\bibfnamefont {Ralph}\ \bibnamefont {Rieger}}, \bibinfo {author}
  {\bibfnamefont {Xinliang}\ \bibnamefont {Feng}}, \bibinfo {author}
  {\bibfnamefont {Klaus}\ \bibnamefont {Müllen}}, \ and\ \bibinfo {author}
  {\bibfnamefont {Roman}\ \bibnamefont {Fasel}},\ }\bibfield  {title} {\bibinfo
  {title} {{Porous graphenes: two-dimensional polymer synthesis with atomic
  precision}},\ }\href {\doibase 10.1039/B915190G} {\bibfield  {journal}
  {\bibinfo  {journal} {Chem. Commun.}\ }\textbf {\bibinfo {volume} {45}},\
  \bibinfo {pages} {6919--6921} (\bibinfo {year} {2009})}\BibitemShut {NoStop}%
\bibitem [{\citenamefont {Villalobos}\ \emph {et~al.}(2020)\citenamefont
  {Villalobos}, \citenamefont {Vahdat}, \citenamefont {Dakhchoune},
  \citenamefont {Nadizadeh}, \citenamefont {Mensi}, \citenamefont {Oveisi},
  \citenamefont {Campi}, \citenamefont {Marzari},\ and\ \citenamefont
  {Agrawal}}]{villalobos2020large}%
  \BibitemOpen
  \bibfield  {author} {\bibinfo {author} {\bibfnamefont {Luis~Francisco}\
  \bibnamefont {Villalobos}}, \bibinfo {author} {\bibfnamefont
  {Mohammad~Tohidi}\ \bibnamefont {Vahdat}}, \bibinfo {author} {\bibfnamefont
  {Mostapha}\ \bibnamefont {Dakhchoune}}, \bibinfo {author} {\bibfnamefont
  {Zahra}\ \bibnamefont {Nadizadeh}}, \bibinfo {author} {\bibfnamefont
  {Mounir}\ \bibnamefont {Mensi}}, \bibinfo {author} {\bibfnamefont {Emad}\
  \bibnamefont {Oveisi}}, \bibinfo {author} {\bibfnamefont {Davide}\
  \bibnamefont {Campi}}, \bibinfo {author} {\bibfnamefont {Nicola}\
  \bibnamefont {Marzari}}, \ and\ \bibinfo {author} {\bibfnamefont
  {Kumar~Varoon}\ \bibnamefont {Agrawal}},\ }\bibfield  {title} {\bibinfo
  {title} {{Large-scale synthesis of crystalline g-C$_3$N$_4$ nanosheets and
  high-temperature H$_2$ sieving from assembled films}},\ }\href
  {https://www.science.org/doi/abs/10.1126/sciadv.aay9851} {\bibfield
  {journal} {\bibinfo  {journal} {Sci. Adv.}\ }\textbf {\bibinfo {volume}
  {6}},\ \bibinfo {pages} {eaay9851} (\bibinfo {year} {2020})}\BibitemShut
  {NoStop}%
\bibitem [{\citenamefont {Tan}\ \emph {et~al.}(2016)\citenamefont {Tan},
  \citenamefont {Chen}, \citenamefont {Xia}, \citenamefont {Liu},\ and\
  \citenamefont {Xu}}]{tan2016parent}%
  \BibitemOpen
  \bibfield  {author} {\bibinfo {author} {\bibfnamefont {Qitao}\ \bibnamefont
  {Tan}}, \bibinfo {author} {\bibfnamefont {Huanhuan}\ \bibnamefont {Chen}},
  \bibinfo {author} {\bibfnamefont {Huaida}\ \bibnamefont {Xia}}, \bibinfo
  {author} {\bibfnamefont {Bingxin}\ \bibnamefont {Liu}}, \ and\ \bibinfo
  {author} {\bibfnamefont {Bin}\ \bibnamefont {Xu}},\ }\bibfield  {title}
  {\bibinfo {title} {{Parent and trisubstituted triazacoronenes: synthesis,
  crystal structure and physicochemical properties}},\ }\href
  {https://doi.org/10.1039/C5CC08853D} {\bibfield  {journal} {\bibinfo
  {journal} {Chem. Commun.}\ }\textbf {\bibinfo {volume} {52}},\ \bibinfo
  {pages} {537--540} (\bibinfo {year} {2016})}\BibitemShut {NoStop}%
\bibitem [{\citenamefont {Wang}\ \emph {et~al.}(2015)\citenamefont {Wang},
  \citenamefont {Dai}, \citenamefont {Li}, \citenamefont {Yang}, \citenamefont
  {Srolovitz},\ and\ \citenamefont {Zheng}}]{wang2015measurement}%
  \BibitemOpen
  \bibfield  {author} {\bibinfo {author} {\bibfnamefont {Wen}\ \bibnamefont
  {Wang}}, \bibinfo {author} {\bibfnamefont {Shuyang}\ \bibnamefont {Dai}},
  \bibinfo {author} {\bibfnamefont {Xide}\ \bibnamefont {Li}}, \bibinfo
  {author} {\bibfnamefont {Jiarui}\ \bibnamefont {Yang}}, \bibinfo {author}
  {\bibfnamefont {David~J}\ \bibnamefont {Srolovitz}}, \ and\ \bibinfo {author}
  {\bibfnamefont {Quanshui}\ \bibnamefont {Zheng}},\ }\bibfield  {title}
  {\bibinfo {title} {{Measurement of the cleavage energy of graphite}},\ }\href
  {https://doi.org/10.1038/ncomms8853} {\bibfield  {journal} {\bibinfo
  {journal} {Nat. Commun.}\ }\textbf {\bibinfo {volume} {6}},\ \bibinfo {pages}
  {7853} (\bibinfo {year} {2015})}\BibitemShut {NoStop}%
\bibitem [{\citenamefont {Meyer}\ \emph {et~al.}(2007)\citenamefont {Meyer},
  \citenamefont {Geim}, \citenamefont {Katsnelson}, \citenamefont {Novoselov},
  \citenamefont {Booth},\ and\ \citenamefont {Roth}}]{meyer2007structure}%
  \BibitemOpen
  \bibfield  {author} {\bibinfo {author} {\bibfnamefont {Jannik~C}\
  \bibnamefont {Meyer}}, \bibinfo {author} {\bibfnamefont {Andre~K}\
  \bibnamefont {Geim}}, \bibinfo {author} {\bibfnamefont {Mikhail~I}\
  \bibnamefont {Katsnelson}}, \bibinfo {author} {\bibfnamefont {Konstantin~S}\
  \bibnamefont {Novoselov}}, \bibinfo {author} {\bibfnamefont {Tim~J}\
  \bibnamefont {Booth}}, \ and\ \bibinfo {author} {\bibfnamefont {Siegmar}\
  \bibnamefont {Roth}},\ }\bibfield  {title} {\bibinfo {title} {{The structure
  of suspended graphene sheets}},\ }\href
  {https://www.nature.com/articles/nature05545} {\bibfield  {journal} {\bibinfo
   {journal} {Nature}\ }\textbf {\bibinfo {volume} {446}},\ \bibinfo {pages}
  {60--63} (\bibinfo {year} {2007})}\BibitemShut {NoStop}%
\end{thebibliography}%
 \end{document}